\documentclass[%
 reprint,
 amsmath,amssymb,
 aps,
]{revtex4-2}

\usepackage{graphicx}
\usepackage{dcolumn}
\usepackage{bm}
\usepackage{hyperref}
\usepackage{xspace}
\usepackage{xifthen}
\usepackage{xcolor}

\newcommand{\pdf}{p.d.f\mbox{.}\xspace}
\newcommand{\pdfs}{p.d.f\mbox{.s}\xspace}
\newcommand{\sweights}{\emph{sWeights}\xspace}
\newcommand{\dd}[1]{\text{d}#1}
\newcommand{\intd}[1]{\int\!\dd{#1}}
\newcommand{\de}[1]{\,\dd{#1}}
\newcommand{\eq}[1]{Eq.\,\ref{eq:#1}}
\newcommand{\fig}[1]{Fig.\,\ref{fig:#1}}
\DeclareMathOperator{\ex}{E}
\DeclareMathOperator{\var}{Var}

\newcommand{\ie}{\emph{i.e.\ }}
\newcommand{\eg}{\emph{e.g.\ }}
\newcommand{\lnL}{\ln \!\mathcal{L}}
\newcommand{\eps}{\epsilon}
\newcommand{\we}[1][]{%
\ifthenelse{\isempty{#1}}{w_\eps}{w_{\eps,#1}}%
}

\begin{document}

\preprint{arxiv:????-????}

\title{Custom Orthogonal Weight functions (COWs) for Event Classification}

\author{Hans Dembinski}
 \email{hans.dembinski@tu-dortmund.de}
 \affiliation{TU Dortmund, Germany}
\author{Matthew Kenzie}%
 \email{matthew.kenzie@cern.ch}%
 \affiliation{University of Warwick, United Kingdom}
\author{Christoph Langenbruch}%
 \email{christoph.langenbruch@cern.ch}
 \affiliation{RWTH Aachen, Germany}
\author{Michael Schmelling}%
 \email{michael.schmelling@mpi-hd.mpg.de}
 \affiliation{Max Planck Institute for Nuclear Physics, Heidelberg, Germany}

\date{\today}

\begin{abstract}
A common problem in data analysis is the separation of signal and background. We revisit and generalise the so-called \sweights method, which allows one to calculate an empirical estimate of the signal density of a control variable using a fit of a mixed signal and background model to a discriminating variable. We show that \sweights are a special case of a larger class of Custom Orthogonal Weight functions (COWs), which can be applied to a more general class of problems in which the discriminating and control variables are not necessarily independent and still achieve close to optimal performance. We also investigate the properties of parameters estimated from fits of statistical models to \sweights and provide closed formulas for the asymptotic covariance matrix of the fitted parameters. To illustrate our findings, we discuss several practical applications of these techniques.
\end{abstract}

\maketitle


\section{Introduction}

This article takes a fresh look at the \sweights (or \textit{sPlot}) formalism discussed by Barlow \cite{Barlow:1986ek} and popularised more recently by Pivk and Le Diberder~\cite{Pivk:2004ty}. The \sweights method is used to infer properties of a signal distribution in a mixed data set containing signal and background events. The signal distribution is extracted non-parametrically by applying weights to individual events. Inference is then done on the weighted data set. The method is applicable, when individual points from the data distribution consist of a discriminating variable(s), here called $m$, and one or more statistically independent control variables, here called $t$, where $m$ and $t$ can both be vectors and of different dimensions. By fitting parametric models to the signal and background in the discriminating variable(s) $m$, one can calculate the weight distribution that represents the signal density in the control variable(s) $t$. The advantage of this method, compared to a fully parametric fit to the $(m, t)$ distribution, is that one avoids the need to parameterise the background density in the control variable(s) $t$, which is often   challenging.

In Sec.~\ref{sec:sweights} we re-derive the \sweights method from the starting point of orthonormal functions. We show several ways of calculating the weights and compare their trade-offs, and emphasise that \sweights can easily be computed without some of the restrictions seen previously.

In Sec.~\ref{sec:cows} we then discuss a generalisation of the \sweights method called ``Custom Orthogonal Weight functions" (COWs). COWs relax some of the requirements of the \sweights formalism and can be applied to a larger class of problems than traditional \sweights, at a small loss in precision.

In Sec.~\ref{sec:uncerts} we then discuss the properties of estimates obtained when fitting models to weighted data. We give an asymptotically correct formula for the covariance matrix of the parameters obtained from such a fit.

Finally in Sec.~\ref{sec:application} we perform a variety of studies on simulated Monte Carlo which deploy \sweights and COWs on various applications and show comparisons of their performance.

\section{sWeights as orthonormal functions}
\label{sec:sweights}

To compute the weights for the signal distribution in the control variable $t$, we use a discriminant variable $m$ (often the invariant mass of some particle's decay products). The signal and background density only need to be parameterised in the discriminant variable $m$. The variables $m$ and $t$ must be statistically independent in the classic \sweights formalism, so that the respective \pdf{}s of the variables factorise. In other words, we assume that the total \pdf has the following form
\begin{equation}\label{eq:fmt}
f(m, t) = z\, g_s(m) \, h_s(t) + (1 - z) \, g_b(m) \, h_b(t),
\end{equation}
where $z$ is the signal fraction, $g_s(m)$ and $h_s(t)$ are the signal \pdf{}s in the discriminating and control variables, respectively, and $g_b(m)$ and $h_b(t)$, the corresponding background \pdf{}s. The \sweights method allows one to obtain an asymptotically efficient non-parametric estimate of $z \, h_s(t)$ while only requiring parametric models for $g_s(m)$ and $g_b(m)$.

We stress that the \sweights method is only applicable when the \pdf{}s in $m$ and $t$ factorise for both the signal and the background, which is conditional on their independence. Independence is a stronger condition than lack of correlation. Therefore, tests which demonstrate a lack of correlation between $m$ and $t$ provide necessary, but not sufficient, evidence for the applicability of the \sweights method. We come back to proper tests of independence in Sec.~\ref{sec:application}.

\subsection{Construction of an optimal weight function}
\label{sec:sweights:optimal}

We postulate that a weight function, $w_s(m)$, exists which extracts the signal component, $z \, h_s(t)$, when $f(m, t)$ is multiplied by it and integrated over $m$:
\begin{multline}\label{eq:partial_expectation}
z \, h_s(t) \overset != \intd m\, w_s(m)\, f(m, t) = \\
\intd m\, w_s(m)\, \big[z\, g_s(m) \, h_s(t) + (1 - z) \, g_b(m) \, h_b(t) \big] \\
= z \, h_s(t) \intd m \, w_s(m) \, g_s(m) \\
 + (1 - z) \, h_b(t) \intd m \, w_s(m) \, g_b(m).
\end{multline}
The left and the right-hand sides of \eq{partial_expectation} are equal in general only if the following conditions hold:
\begin{align}
\intd m \, w_s(m) \, g_s(m) &= 1 \label{eq:normal} \\
\intd m \, w_s(m) \, g_b(m) &= 0. \label{eq:orthogonal}
\end{align}
If we regard $\intd m \, \phi(m) \, \psi(m)$ as the inner product of a vector space over functions, then these conditions define $w_s(m)$ as the vector orthogonal to $g_b(m)$ and normal to $g_s(m)$. In other words, $w_s(m)$ is an orthonormal function in this space.

Since the vector space over $m$ is infinite-dimensional, there are infinitely many orthonormal functions $w_s(m)$ that satisfy these conditions. For example, the classic sideband subtraction method can be regarded as a special case where $w_s(m)$ is a piece-wise constant function which is positive in the signal region and negative in the background region.

In order to obtain a unique solution for $w_s(m)$ we can chose to minimise its variance. Since $f(m, t)$ factorises and $w_s(m)$ is only a function of $m$, we can obtain all information about $w_s$ from the density $g(m)$, computed by integrating \eq{fmt} over $t$,
\begin{equation}
g(m) = \intd t \, f(m, t) = z \, g_s(m) + (1 - z) \, g_b(m).
\end{equation}
The expectation of $w_s$ over $g(m)$ is
\begin{equation}\label{eq:exp_w_s}
\ex[w_s] = \int w_s(m) \, g(m) \de m = z,
\end{equation}
and the variance of $w_s$ over $g(m)$ is given by
\begin{equation}\label{eq:var_w_s}
\var(w_s) = \ex[w_s^2] - \ex[w_s]^2 = \int w_s(m)^2 \, g(m) \de m - z^2.
\end{equation}
Minimising the variance $\var(w_s)$ guarantees that the sample estimate $\hat z = 1/N \, \sum_i^N \hat w_s(m_i)$ asymptotically has minimum variance. As a byproduct, this choice also produces minimum variance for the estimated background fraction $(1 - \hat z)$, and generally smooth functions, $w_s(m)$, since oscillating solutions have larger variance.

To find the function $w_s(m)$ which minimises $\var(w_s)$, we have to solve a constrained minimisation problem. The solution, computed in Appendix~\ref{sec:constrained_opt}, is
\begin{equation}\label{eq:w_s}
w_s(m) = \frac{\alpha_s \, g_s(m) + \alpha_b \, g_b(m)}{g(m)}.
\end{equation}
The constants $\alpha_{s,b}$ are obtained by inserting \eq{w_s} into \eq{normal} and \eq{orthogonal} and solving the resulting system of linear equations. Before we continue with that, we note that the signal component plays no special role in the derivation so far. We could have equally postulated a weight function $w_b(m)$ to extract the background, which leads to the conditions
\begin{align}
\intd m \, w_b(m) \, g_s(m) &= 0 \\
\intd m \, w_b(m) \, g_b(m) &= 1,
\end{align}
and
\begin{equation}\label{eq:w_b}
w_b(m) = \frac{\beta_s \, g_s(m) + \beta_b \, g_b(m)}{g(m)}.
\end{equation}
The coefficients $\alpha_x$ and $\beta_x$ with $x \in \{s, b\}$ can be computed by solving
\begin{equation}\label{eq:w_matrix}
\underbrace{\begin{pmatrix}
W_{ss} & W_{sb} \\
W_{sb} & W_{bb}
\end{pmatrix}}_{\bm W}
\cdot
\underbrace{\begin{pmatrix}
\alpha_s & \beta_s \\
\alpha_b & \beta_b
\end{pmatrix}}_{\bm A}
=
\begin{pmatrix}
1 & 0 \\
0 & 1
\end{pmatrix},
\end{equation}
with
\begin{equation}\label{eq:wxy}
W_{xy} = \intd m \, \frac{g_x(m) \, g_y(m)}{g(m)}.
\end{equation}
In other words, the matrix $\bm A$, formed by the coefficients to compute $w_s(m)$ and $w_b(m)$, is the inverse of the symmetric positive-definite $\bm W$ matrix.

With Cramer's rule, we get
\begin{align}
\alpha_s &= \frac{W_{bb}}{W_{ss} W_{bb} - W_{sb}^2} &
\alpha_b &= \frac{-W_{sb}}{W_{ss} W_{bb} - W_{sb}^2}, \\
\beta_s &= \frac{-W_{sb}}{W_{ss} W_{bb} - W_{sb}^2} &
\beta_b &= \frac{W_{ss}}{W_{ss} W_{bb} - W_{sb}^2},
\end{align}
One can further replace $g(m)$ in the denominator of \eq{w_s} (or \eq{w_b}) by inserting \eq{w_s} into \eq{exp_w_s} to find that $z = \alpha_s + \alpha_b$, and similarly one finds $1 - z = \beta_s + \beta_b$. With these ingredients, we obtain the final equations
\begin{align}\label{eq:w_x_final}
w_s(m) &= \frac{W_{bb} \, g_s(m) - W_{sb} \, g_b(m)}{(W_{bb} \!-\! W_{sb}) \, g_s(m) + (W_{ss} \!-\! W_{sb}) g_b(m)}, \\
w_b(m) &= \frac{W_{ss} \, g_b(m) - W_{sb} \, g_s(m)}{(W_{bb} \!-\! W_{sb}) \, g_s(m) + (W_{ss} \!-\! W_{sb}) g_b(m)}.
\end{align}
In summary, to obtain $w_s(m)$ or $w_b(m)$ one has to compute the matrix elements $W_{ss}, W_{sb}, W_{bb}$, which depend only on $g_{s,b}(m)$ and (implicitly) $z$.

\subsection{Application to finite samples}
\label{sec:app_fin_samps}

The calculations so far were carried out for the true \pdf{}s, $g_{s,b}(m)$, and true signal fraction, $z$, on which the matrix elements $W_{xy}$ depend. In practice, these need to be replaced by sample estimates $\hat g_{s,b}(m)$ and $\hat z$, typically obtained from a maximum-likelihood fit, although any kind of estimation can be used. The plug-in estimate \cite{Efron:1986hys} of \eq{w_x_final} is
\begin{equation}\label{eq:hat_w_s}
\hat w_s(m) =
\frac
{\widehat W_{bb} \, \hat g_s(m) - \widehat W_{sb} \, \hat g_b(m)}
{(\widehat W_{bb} \!-\! \widehat W_{sb}) \, \hat g_s(m) + (\widehat W_{ss} \!-\! \widehat W_{sb}) \, \hat g_b(m)}.
\end{equation}
For the computation of the estimates $\widehat W_{xy}$ with $x\in \{s,b\}$ we face a choice between two possibilities.

\begin{itemize}
\item \textit{Variant A:} We replace the true quantities in \eq{wxy} with their plug-in estimates and compute the integral analytically or numerically,
\begin{equation}\label{eq:wxya}
\widehat W^{A}_{xy} = \intd m \, \frac{\hat g_x(m) \, \hat g_y(m)}{\hat z \, \hat g_s(m) + (1 - \hat z) \, \hat g_b(m)}.
\end{equation}

\item \textit{Variant B}: We additionally replace the integral with a sum over the observations in the data sample. We note that an integral over a function $\phi(m)$ can be written as an expectation value over the \pdf $g(m)$ (assuming that the expectation exists),
\begin{equation}
\intd m \, \phi(m) = \intd m \, g(m) \frac{\phi(m)}{g(m)} = \ex[\phi(m) / g(m)].
\end{equation}
In a finite sample, the arithmetic mean is an unbiased estimate of the expectation due to the law of large numbers,
\begin{equation}\label{eq:exp2mean}
\frac{1}{N} \sum_i \frac{\phi(m_i)}{g(m_i)} \longrightarrow \ex[\phi(m) / g(m)],
\end{equation}
where $m_i$ is the $i$-th observed value of $m$ and $N$ is the sample size. Applying this replacement to \eq{wxy} yields
\begin{equation}\label{eq:wxyb}
\widehat W^{B}_{xy} = \frac{1}{N} \sum_i \frac{\hat g_x(m_i) \, \hat g_y(m_i)}{\big(\hat z \hat g_s(m_i) + (1 - \hat z) \hat g_b(m_i)\big)^2}.
\end{equation}
\end{itemize}
\textit{Variant B} has several attractive properties which make it the recommended method. The computation is straight-forward from the fitted estimates $\hat z$ and $\hat g_{s,b}(m)$ and the data sample. The additional complexity of computing an integral (possibly numerically) is avoided. Furthermore, this choice is guaranteed to exactly reproduce the previously fitted signal yield $\hat N_s = N \hat z$ when the \sweights are summed:
\begin{equation}\label{eq:sum_w_s_equal_nz}
\sum_i \hat w_s(m_i) = N \hat z = \hat N_s,
\end{equation}
where $\hat w_s(m)$ is the estimate of $w_s(m)$ computed from $\widehat{W}^B_{xy}$. The proof for this is provided in Appendix \ref{sec:sum_w_s_equal_nz}.

In other words, \textit{Variant B} produces self-consistent estimates $\hat w_s$ for the sample at hand. This is not exactly true in general for \textit{Variant A}. The matrix elements $\widehat{W}^{A}_{xy}$ are numerically close to the elements $\widehat{W}^{B}_{xy}$ but differ. We consider the self-consistency of \eq{sum_w_s_equal_nz} important: \sweights are computed from a fitted estimate $\hat z$, and so they should reproduce that estimate exactly.

\subsection{Connection to extended maximum-likelihood fit}
\label{sec:maxlh}

There is a curious connection between \eq{wxyb} and the results of an extended maximum-likelihood fit in which $\hat g_s$ and $\hat g_b$ are fixed to their maximum-likelihood estimates and the respective signal and background yields, $N_s$ and $N_b$, are regarded as independent variables. In such a fit, one maximises the extended log-likelihood function ~\cite{Barlow:1990vc} which is without constant terms
\begin{equation}\label{eq:mlex}
\lnL(N_s, N_b)
= -(N_s + N_b) +
\sum_i \ln [N_s \, \hat g_s(m_i) + N_b \, \hat g_b(m_i)].
\end{equation}
The extremum is determined by solving the score functions
\begin{equation}\label{eq:mlscore}
\frac{\partial \lnL}{\partial N_x} = -1 + \sum_i \frac{\hat g_x(m_i)}{N_s \, \hat g_s(m_i) + N_b \, \hat g_b(m_i)} \overset{!}= 0,
\end{equation}
with $x \in \{s,b\}$. The maximum-likelihood estimates obtained from these score functions are $\hat N_s = N \hat z$ and $\hat N_b = N (1 - \hat z)$, where $\hat z$ is the estimated signal fraction as before. The elements of the Hessian matrix, of second derivatives of the log-likelihood function, are given by
\begin{equation}\label{eq:hessian}
\frac{\partial^2 \lnL}{\partial N_x \, \partial N_y} = -\sum_i \frac{\hat g_x(m_i) \, \hat g_y(m_i)}{\big( N_s \, \hat g_s(m_i) + N_b \, \hat g_b(m_i) \big)^2}.
\end{equation}
We note the similarity between \eq{hessian} and \eq{wxyb} and evaluate the second derivative at the maximum of $\lnL$ to find
\begin{multline}\label{eq:wxyc}
-\frac{\partial^2 \lnL}{\partial N_x \, \partial N_y}\bigg|_{N_s = N \hat z,\, N_b = N (1-\hat z)} \\
= \sum_i \frac{\hat g_x(m_i) \, \hat g_y(m_i)}{\big( N \, \hat z \, \hat g_s(m_i) + N \, (1-\hat z) \, \hat g_b(m_i) \big)^2} \\
= \frac{1}{N} \widehat W^B_{xy}.
\end{multline}

This shows another opportunity to compute estimates of $W_{xy}$, since the second derivatives of the log-likelihood are routinely computed (for example, by the program MINUIT) as part of the fit for $\hat N_s$, $\hat N_b$, and the shape parameters $\bm \theta_{s,b}$ of $\hat g_{s,b}(m; \bm \theta_{s, b})$, and are therefore readily available. The covariance matrix $\mathbf C$ returned by such a fitting program is the negative inverse of the Hessian,
\begin{equation}\label{eq:mlcov}
\mathbf{C}^{-1} = -
\begin{pmatrix}
\frac{\partial^2 \lnL}{\partial N_s^2} & \frac{\partial^2 \lnL}{\partial N_s \partial N_b} & \dots \\
\frac{\partial^2 \lnL}{\partial N_s \partial N_b} & \frac{\partial^2 \lnL}{\partial N_b^2} & \dots \\
\vdots & \vdots & \ddots \\
\end{pmatrix}.
\end{equation}
The dotted parts of the matrix correspond to derivatives that contain one or two shape parameters of $\bm \theta_{s,b}$.

Thus one can use \emph{Variant C} to compute the elements of $\widehat W^C_{xy}$ which consists of the following steps:
\begin{itemize}
\item Invert the covariance matrix $\mathbf C$ of the fit of yields $N_{s,b}$ and shape parameters $\bm \theta_{s,b}$.
\item Isolate the $2 \times 2$ sub-matrix of the Hessian which contains the derivatives with respect to the yields $N_{s,b}$.
\item Use \eq{wxyc} on these matrix elements to obtain $\widehat W^C_{xy}$.
\end{itemize}
It would be incorrect to switch steps 1 and 2, \ie isolate the $2 \times 2$ sub-matrix of $\mathbf C$ that contains the yields and invert it, because this does not restore the derivatives.

A close alternative is to do a second fit which leaves only the yields free while keeping shape parameters fixed. In this case, the covariance matrix computed by MINUIT can be scaled to yield an estimate of the coefficient matrix from \eq{w_matrix}:
\begin{equation}\label{eq:dir_cov}
\begin{pmatrix}
\hat \alpha_s & \hat \beta_s \\
\hat \alpha_b & \hat \beta_b
\end{pmatrix}
= \frac{1}{N}
\begin{pmatrix}
C_{ss} & C_{sb} \\
C_{sb} & C_{bb}
\end{pmatrix}.
\end{equation}

If the Hessian matrix was actually calculated with \eq{hessian}, \emph{Variant B} and \emph{C} would give identical results. In practice however, the second derivatives in \eq{mlcov} are usually computed only approximately by numerical differentiation of \eq{mlex}. The accuracy of numerical differentiation is several orders below the machine precision. This means that \emph{Variant C} produces a less accurate estimate than \emph{Variant B} and that \eq{sum_w_s_equal_nz} only holds approximately for \emph{Variant C}. In conclusion, \emph{Variant B} is recommended over \emph{Variant C}, since the computation is inexpensive and the result more accurate.

\newcommand{\vev}[1]{{\rm E}\left[#1\right]}

\section{Custom orthogonal weight functions}
\label{sec:cows}
The discussion so far has focused on the restricted case where the \pdf is a mixture of two components that each factorise in both the discriminant and control variables. We now generalise to an arbitrary number of factorising components, and also allow for a non-factorising function of frequency weights, $\eps(m,t)$, which in practical applications is often identified with an efficiency function. The total \pdf for the observed data then becomes
\begin{multline}
    \rho(m,t) = \frac{1}{D} \eps(m,t) f(m,t) \\
    \mbox{with}\quad
    D = \int dm\,dt\,\eps(m,t)\,f(m,t) \;.
\end{multline}
The normalisation term, $D$, ensures that the observed density, $\rho(m,t)$, is properly normalised. The true density of interest is
\begin{multline}
    f(m,t) = \sum_{k=0}^n z_k g_k(m) h_k(t)
    \quad\mbox{with}\quad
    \sum_{k=0}^n z_k = 1 \;.
\end{multline}
The Kolmogorov–Arnold representation theorem~\cite{Kolmogorov1957,Arnold2009} ensures that a finite sum of terms on the right-hand side can represent any two-dimensional function $f(m,t)$. For practical applications it is beneficial if the expansion requires only a few terms, which can be achieved with $g_k(m)$ and $h_k(t)$ suitably chosen for the specific case. For a given expansion we will assume that the first $s$ terms pertain to the signal density while the others describe the background, i.e.
\begin{multline}
\label{eq:fmt_expansion}
   f(m,t) = \underbrace{\sum_{k=0}^{s-1} z_k\,g_k(m)\,h_k(t)}_{\mbox{\scriptsize signal}}
   + \underbrace{\sum_{k=s}^n z_k\,g_k(m)\,h_k(t)}_{\mbox{\scriptsize background}} \;.
\end{multline}
If there are multiple terms, in either the signal or background part, that do not contain either identical $g_k(m)$ or $h_k(t)$ components, then the respective \pdfs are non-factorising.

Generalising the insights obtained when identifying the \sweights as orthogonal functions (see Sec.~\ref{sec:sweights:optimal}), it is easy to show that any single function $h_k(t)$ in $f(m,t)$ can be isolated by a weight function
\begin{multline}
\label{eq:cow}
    w_k(m) = \sum_{l=0}^n \frac{A_{kl}\,g_l(m)}{I(m)} \\
    \mbox{with,}\quad
    A^{-1}_{kl} = W_{kl} = \int dm\,\frac{g_k(m)\,g_l(m)}{I(m)} \;.
\end{multline}
Here $I(m)$ is an arbitrary function (which we hereafter refer to as the ``variance function"), that is only required to be non-zero in the considered range of $m$, and $A_{kl}$ is akin to the $\alpha$, $\beta$ matrix of \eq{w_matrix}. It follows that
\begin{multline}
    \sum_{i=0}^n A_{ki}\,W_{ij} = \delta_{kj}
    \;\;\;\mbox{and}\;
    \int dm\, w_k(m)\,g_l(m) = \delta_{kl}\;.
\end{multline}
The weight functions, $w_k(m)$, are orthonormal to the \pdfs, $g_k(m)$, in the discriminant variable, subject to the weight $I(m)$. For $I(m)=1$, $m\in[-1,+1]$ and $g_k(m)=m^k$, the $w_k(m)$ would be the Legendre polynomials.
For $I(m)=\sqrt{1-m^2}$, one would obtain the Chebychev polynomials. For a particular problem, the basis functions, $g_k(m)$, and the choice of the weight function, $I(m)$, thus determine a set of Custom Orthogonal Weights functions (COWs).

When considering a non-uniform efficiency, $\eps(m,t)\neq 1$, the
appropriate weight to apply to the data, in order to extract the density $h_k(t)$, is $w_k(m)/\eps(m,t)$.
For a particular bin in the control variable, $\Delta t$, the expectation value of this weight is
\begin{equation}
\label{eq:zD}
    \text{E}_k = \vev{\frac{w_k(m)}{\eps(m,t)}}_{\Delta t}
    = \frac{z_k}{D}\int_{\Delta t} dt\,h_k(t),
\end{equation}
\ie an unbiased estimate for the integral of the efficiency corrected density, $h_k(t)$, over the bin $\Delta t$. This holds for \emph{any} choice $I(m)$. The weights that project out the entire signal or background component are given by
\begin{equation}
    w_s =\sum_{k=0}^{s-1} \frac{w_k(m)}{\eps(m,t)}
    \quad\mbox{and}\quad
    w_b = \sum_{k=s}^n \frac{w_k(m)}{\eps(m,t)} \;.
\end{equation}
Integrating Eq.~\eqref{eq:zD} over all $t$ one sees that every expectation value, $\text{E}_k$, is proportional to $z_k$. Therefore, an estimate of $\hat{z}_k$ can be obtained from the corresponding sample average of $w_k(m)/\eps(m,t)$, with $1/D$ estimated by the sample average of $1/\eps(m,t)$.

Special properties hold when $I(m)$ is a linear combination of the basis functions, $g_k(m)$.  As proven in Appendix~\ref{sec:sum_weights_unity}, for arbitrary constants $a_k$ (where $k\in[0,n]$), one finds
\begin{equation}
    \sum_{k=0}^n w_k(m) = 1
    \quad\mbox{when}\quad
    I(m) = \sum_{k=0}^n a_k\,g_k(m) \;,
\end{equation}
\ie every event contributes with a total weight of unity to the possible states $k$. One corollary of this result is that for every measured $m_i$, the COWs, $w_k(m_i)$, sum to unity when one of the $g_k(m)$ is constant. Another consequence is that with an increasing number of terms the sum $\sum_k w_k(m)$ will converge towards unity for any function $I(m)$, since a linear combination of sufficiently many basis functions $g_k(m)$ always allows for a good approximation of $I(m)$.

It remains to select the weight function $I(m)$. While $I(m)=1$ may be a reasonable default, it certainly will not be optimal. Here we consider two options to choose a better weight function $I(m)$, such that
\begin{enumerate}
    \item the variances of the $\hat{z}_k$ are minimal,
    \item the $\hat{z}_k$ are the Maximum Likelihood estimates.
\end{enumerate}

As shown in Appendix~\ref{sec:app:min_var_zhat}, requirement (1) leads to
\begin{equation}
    \label{eq:cow_estimate}
   I(m) = q(m)
   \quad\mbox{with}\quad
   q(m) = \int dt\,\frac{\rho(m,t)}{\eps^2(m,t)} \;.
\end{equation}
Numerically, $q(m)$ can be obtained from a histogram of the $1/\eps^2(m,t)$ weighted $m$-distribution or a suitable parameterisation thereof. For the construction of the COWs the exact form of $I(m)$ is uncritical, therefore a histogram approximation will usually be good enough. The extreme case of a single-bin histogram is equivalent to $I(m)=1$. Asymptotically a sufficiently fine-binned histogram will be arbitrarily close to the ideal $q(m)$.

Appendix~\ref{sec:app:max_like_zhat} shows that the alternative requirement (2) leads to
\begin{equation}
    \label{eq:cow_norm_func}
    I(m) = \sum_{l=0}^n \hat{z}_k\,g_k(m) \;,
\end{equation}
where the $\hat{z}_k$ are estimates for the true fractions $z_k$ obtained from an $1/\eps(m,t)$ weighted unbinned Maximum Likelihood fit. For $\eps(m,t)=1$ this \emph{is} the \sweights solution. Numerically the $\hat{z}_k$ can be determined iteratively, starting with \eg $\hat{z}_k=1/n$ and updating the values using sample averages of $w_k(m)/\eps(m,t)$, based on the resulting weight functions, $w_k(m)$. Since any initial choice for $I(m)$ yields unbiased estimates, $\hat{z}_k$, the iteration converges quickly. Numerical studies indicate that $n$ steps, where $n$ is the number of coefficients, are usually sufficient. In the case of non-uniform efficiencies, $\eps(m,t)$, the weight function, $I(m)$, from the Maximum Likelihood criterion is different from $q(m)$. The fact that $q(m)$ was derived from the requirement of minimum variance illustrates the known result that weighted Maximum Likelihood estimates are in general not efficient.

It is interesting to compare the two options discussed for $I(m)$ in the case of uniform efficiency weights, $\eps(m,t)=1$.
In this case the weight functions are
\begin{equation}
    I^{(1)}(m) = \sum_{l=0}^n z_k\,g_k(m)
    \quad\mbox{and}\quad
    I^{(2)}(m) = \sum_{l=0}^n \hat{z}_k\,g_k(m) \;,
\end{equation}
\ie the \sweights solution, $I^{(2)}(m)$, \emph{is} the Maximum Likelihood estimate of the theoretically optimal weight function, $I^{(1)}(m)$. Asymptotically $I^{(1)}(m)$ and $I^{(2)}(m)$ are the same. For a non-uniform efficiency function this will not generally be true. One also finds that the COWs, $w_k(m)$, determined from \eq{cow}, with $I(m)=I^{(2)}(m)$, satisfy the consistency condition $1/N \sum_{i=1}^N w_k(m)=\hat{z}_k$ found for \sweights when using the respective sample averages for $W_{kl}$.

\subsection{COWs in the Wild}
\label{sec:cows_in_wild}
The previous section covers the general framework regarding COWs. It shows how one can extract a true density, $h_k(t)$, in the control variable, $t$, from efficiency-distorted data by using only \pdfs, $g_k(m)$, in the discriminant variable, $m$.
In the discussion above, these densities, $g_k(m)$, are defined at the truth level. However, for practical applications these are usually unknown, and additional considerations come into play.

If the efficiency function is not sufficiently well known, it may be preferable to first separate signal and background and handle the efficiency corrections in a later step of the analysis. This case is covered in the COWs framework by simply setting $\eps(m,t)=1$. However, one has to keep in mind that even when the true signal density factorises in $m$ and $t$, the efficiency function in general will not, and thus sufficiently many terms in the signal part of the data model are required to account for factorisation-breaking effects. Furthermore, once the signal density, $h_s(t)$, has been determined, the efficiency correction must be done with the signal efficiency projected into just the control variable, $\bar{\eps}(t)$. This can be obtained by averaging $\eps(m,t)$ over $m$ using
\begin{equation}
   \bar{\eps}(t) = \int dm\,\eps(m,t)\,f_s(m,t) \;.
\end{equation}
Here $f_s(m,t)$ denotes the signal part of the true \pdf. If the efficiency function factorises in $m$ and $t$, $\eps(m,t)=\eta(m)\eps(t)$, the $m$ averaged efficiency can be expressed as
\begin{equation}
  \bar{\eps}(t)
  = \eps(t) \left(\int dm\,\frac{g_s(m)}{\eta(m)}\right)^{-1} \;,
\end{equation}
where $g_s(m)$ is the observed signal \pdf in $m$. Normally one will get $\bar{\eps}(t)$ from a Monte Carlo simulation of the signal. One can then also directly apply weights $w_k(m)/\bar{\eps}(t)$ when filling the respective $t$-histogram.
It should be noted that using
$w_k(m)/\eps(m,t)$ as an event-by-event weight instead would be manifestly wrong, since the $m$-dependence in the efficiency factor destroys the orthogonality relations for the COW, and the signal estimate in $t$ becomes polluted by background.

Another use case is a signal component that can be assumed to factorise in $m$ and $t$ on top of a background that may be non factorising. In the above formalism the signal \pdf is then $g_0(m)h_0(t)$, and if one is only interested in projecting out the \pdf, $h_0(t)$, of the signal component, there is additional freedom in the construction of respective COWs. As shown in Appendix~\ref{sec:app:sig_density}, in this case not only arbitrary non-zero weight functions $I(m)$ can be used, but also the assumed signal density can be chosen freely as long as it is not a linear combination of the background \pdfs $g_k(m)$ (where $k=1,\ldots,n$). This may at first glance seem surprising, but just reflects the fact that in order to remove the background in the control variable, $t$, knowledge of the signal shape in the discriminant variable, $m$, is of secondary importance. However, a good description of the background under the signal is crucial.

To construct a signal-only COW, $w_k$, according to \eq{cow} one requires an input model for the signal density, $p(m)$, a set of background \pdfs $g_k(m)$ (where $k=1,\ldots,n$) and a weight function, $I(m)$. Here, $p(m)$ and $g_k(m)$ must be normalised, but the normalisation of $I(m)$ can be arbitrary. The requirement that $w(m)$ cancels all background contributions is
\begin{equation}
    \int dm\,w(m)\,g_k(m) = 0 \quad\mbox{for}\quad k=1,\ldots\,n \;.
\end{equation}
An additional requirement is needed to fix the normalisation of $w(m)$, which is conveniently chosen as
\begin{equation}
    \int dm\,w(m)\,p(m) = 1 \;.
\end{equation}
For $p(m)=g_0(m)$ this is same condition as before. For the case $\eps(m,t)=1$ one can show that $p(m)=g_0(m)$ and $p(m)=\rho(m)$, where $\rho(m)$ is the observed \pdf in $m$, asymptotically give estimates for $h_0(t)$ that have exactly the same statistical accuracy in terms of the number of equivalent events $N_{\rm eq}= (\sum w_i)^2/\sum w_i^2$. This suggests one should use
\begin{equation}
    p(m) = \int dt\,f(m,t) = \sum_{k=0}^n z_k\,g_k(m) \;,
\end{equation}
the \pdf of the efficiency corrected $m$-distribution. Experimentally it can be estimated from the $1/\eps(m,t)$-weighted $m$ distribution of the data. For the optimal choice of the weight function one finds again $I(m)=q(m)$, with $q(m)$ estimated by the $1/\eps^2(m,t)$-weighted $m$-distribution of the data.

This offers an intriguing possibility to extract and estimate the signal \pdf in the control variable, $h_0(t)$, from a set of $N$ measurements $\{m_i,t_i\}$, $i=1,\ldots,N$. All one needs is a model for the background in $m$, and estimates, e.g. histograms, of $p(m)$ and $q(m)$. Formally the background can always be expanded into a complete set of functions, e.g. polynomials. With the conventions adopted above, a factorising model on the interval $m\in [0,1]$ would be
\begin{equation}
    g_b(m,t) = g_b(m)\,h_b(t)
    = \left(\sum_{k=1}^\infty a_k\,k\,m^{k-1}\right) h_b(t)
\end{equation}
a non-factorising model would be obtained by
\begin{equation}
    g_b(m,t) = \sum_{k=1}^\infty a_k\, k\,m^{k-1}h_k(t) \;.
\end{equation}
For practical applications the above sums have to be truncated. If one imposes factorisation of the background, then estimates $\hat{a}_k$ need to be determined from the data in order to specify the background \pdf. If one allows for factorisation breaking, then all one needs are individual \pdfs $g_k(m)=k\,m^{k-1}$, $p(m)$ in place of the actual signal component $g_0(m)$ and $I(m)=q(m)$ to determine $w(m)=w_0(m)$ according to \eq{cow}. The event-by-event weights $w(m)/\eps(m,t)$ for a histogram in $t$ then produce an asymptotically efficient and unbiased estimate of the signal \pdf $h_0(t)$. At finite statistics the use of estimates from the data for $p(m)$ and $q(m)$ will give rise to a bias of order $1/N$, which is negligible compared to the statistical uncertainties. A formal proof for this is still pending. However, any biases will be small since using \emph{a priori} fixed functions for $p(m)$ and $q(m)$, provides an unbiased estimate of $h_0(t)$, although with less than optimal statistical precision. Systematic uncertainties related to the choice of the background model can be probed by adding terms and checking the stability of the result.

\section{Variance of estimates from weighted data}
\label{sec:uncerts}

Parameter estimation using weighted unbinned data sets
can be performed by maximising the weighted likelihood~\cite{James:2006zz}, which is equivalent to solving the weighted score functions
\begin{align}
\sum_{i} w_i \frac{\partial\ln h_s(t_i;\bm{\theta})}{\partial\theta_k} \overset!= 0, \label{eq:weighted_score}
\end{align}
with \sweights $w_i = w_s(m_i)$ or $w_s'(m_i, t_i)$ and shape parameters $\bm\theta$ of the signal \pdf $h_s(t; \bm\theta)$. The weighted likelihood is not a classic likelihood (product of probabilities) and so the inverse of the Hessian matrix~\cite{James:2006zz} of the weighted likelihood does not asymptotically provide an estimate of the covariance matrix of the parameters. \eq{weighted_score} is an example of an \emph{M-estimator}~\cite{1981rost.book.....H}. 
A complete derivation of the asymptotic covariance matrix for the parameters $\bm\theta$ can be found in the appendix of Ref.~\cite{Langenbruch:2019nwe},
here we only summarise the main findings.

A complication arises due to the fact that the \sweights depend, via \eq{hat_w_s}, on the inverse covariance matrix elements $W_{xy}$, which are usually determined via \eq{wxyb}. The estimates $\widehat W_{xy}$ in turn depend on the estimates of the signal and background yields, $\hat N_s$ and $\hat N_b$, usually determined from an extended maximum likelihood fit. 
Problems of this type are described as \textit{two-step M-estimation} in the statistical literature~\cite{wooldridge2010econometric,neweymcfadden}. 
To account for the fact that the parameters are estimated from the same data sample and are therefore not independent, one has to combine the estimating equations for the parameters of interest with those of the yields and the inverse covariance matrix elements in a single vector.

We construct the quasi-score function $\bm S(\bm \lambda)$,
where ${\bm\lambda} =\{N_s, N_b, \bm\phi, W_{ss}, W_{sb}, W_{bb}, \bm\theta\}$ is the vector of all such parameters, and $\bm\phi$ and $\bm\theta$ are also vectors for the shape parameters in $m$ and $t$, respectively. 
The elements of $\bm{S}$ are given by
\begin{equation}
  {\bm S}(\bm \lambda) =
  \left(\begin{array}{c}
    \partial\lnL(N_s, N_b, \bm\phi) / \partial N_s \\
    \partial\lnL(N_s, N_b, \bm\phi) / \partial N_b \\
    \partial\lnL(N_s, N_b, \bm\phi) / \partial {\phi}_1 \\
    \vdots \\
    \partial\lnL(N_s, N_b, \bm\phi) / \partial {\phi}_n \\
    \psi_{ss}(N_s, N_b, \bm\phi, W_{ss}) \\
    \psi_{sb}(N_s, N_b, \bm\phi, W_{sb}) \\
    \psi_{bb}(N_s, N_b, \bm\phi, W_{bb}) \\
    \xi_1(\bm\phi, W_{ss}, W_{sb}, W_{bb},\bm{\theta}) \\
    \vdots \\
    \xi_p(\bm\phi, W_{ss}, W_{sb}, W_{bb},\bm{\theta}) \end{array}\right) \label{eq:score_i_vector},
\end{equation}
where
\begin{align*}
\frac{\partial\ln{\cal L}}{\partial N_x} &= \sum_i \biggl[\frac{g_x(m_i, \bm \phi)}{N_s \, g_s(m_i, \bm \phi) + N_b \, g_b(m_i, \bm \phi)} - \frac 1 N\biggr] \\
\frac{\partial\ln{\cal L}}{\partial \phi_k} &= \sum_i\frac{N_s \, \partial g_s(m_i, \bm \phi)/\partial \phi_k + N_b \, \partial g_b(m_i, \bm\phi)/ \partial \phi_k}{N_s \, g_s(m_i, \bm \phi) + N_b \, g_b(m_i, \bm \phi)} \\
\psi_{(xy)} &= \sum_i\biggl[\frac{g_x(m_i, \bm\phi) \, g_y(m_i, \bm\phi)}{\big(N_s \, g_s(m_i, \bm\phi) + N_b \, g_b(m_i, \bm\phi)\big)^2} - \frac{W_{xy}}{N}\biggr]\\
\xi_k &=\sum_i w_s(m_i; \bm \phi, W_{ss}, W_{sb}, W_{bb}) \dfrac{\partial\ln h_s(t_i;{\bm\theta})}{\partial \theta_k}
\end{align*}
with $x,y \in \{s, b\}$, $(xy)$ iterating over the three unique combinations $\{ ss, sb, bb \}$, and the shape parameters of $\bm \phi$ and $\bm \theta$ running between $\{1\dots n\}$ and $\{1\dots p\}$, respectively. 
For reference these can be compared to the equivalent expressions in \eq{wxyb} and \eq{mlscore}. One can show that $\ex[\bm{S}(\bm \lambda_0)] = {\bm 0}$, if $\bm{\lambda}_0$ is the vector of true parameter values~\cite{Langenbruch:2019nwe}. Therefore, a consistent estimate $\hat{\bm \lambda}$ can be constructed as the solution to $\bm S(\bm \lambda) \overset!= \bm 0$. We note that the elements of $\bm S(\bm \lambda)$ can be multiplied by arbitrary non-zero constants without changing these results.

The asymptotic covariance of $\bm \lambda$, which includes the parameters of interest $\bm\theta$, is then given by~\cite{10.2307/1912526,van2000asymptotic,davison2003statistical}
\begin{equation}
 {\bm C}_{\bm \lambda} = \ex\left[\frac{\partial\bm S}{\partial\bm{\lambda}^T}\right]^{-1}
\times {\bm C}_{\bm S} \times
\ex\left[\frac{\partial\bm S}{\partial\bm{\lambda}^T}\right]^{-T},
\label{eq:sandwichexpectation}
\end{equation}
where $\partial \bm S / \partial \bm \lambda^T$ is defined as the Jacobian matrix built from the derivatives $\partial S_k / \partial \lambda_\ell$ and $\bm C_{\bm S} = \ex\left[{\bm S} {\bm S}^T\right]$.
We note that the inverse of the Jacobian $\partial\bm{S}/\partial\bm{\lambda}^T$ introduces correlations between the parameter uncertainties. 
In a finite sample, the expectation values in Eq.~\ref{eq:sandwichexpectation} can be estimated from the sample. The estimate for $\ex[\partial \bm S / \partial \bm \lambda^T]$ is $\partial \bm S / \partial \bm \lambda^T|_{\hat \lambda}$, while the elements of the matrix $\widehat {\bm C}_{\bm S}$ are provided in Appendix~\ref{app:sampleestimates}.
In the literature, \eq{sandwichexpectation} is often referred to as the \emph{sandwich estimator}, but in this case the variance of the score is modified because we consider fluctuations in the sample size.

In the case of classic \sweights and when the shapes of $g_s(m)$ and $g_b(m)$ are known,
some simplifications of the expressions in \eq{sandwichexpectation} are possible, as detailed in Ref.~\cite{Langenbruch:2019nwe}.
They result in the following covariance matrix
\begin{equation} \label{eq:cov_corr}
    \widehat{\bm C}_{\bm \theta} = \bm{H}^{-1}\bm{H}^{\prime} \bm{H}^{-T} - \bm{H}^{-1} \bm{}\bm{E}\bm{C}^{\prime}\bm{E}^T \bm{H}^{-T},
\end{equation}
for the parameters of interest $\bm{\theta}$, with
\begin{align*}
H_{k\ell} =& \sum_i \hat w_s(m_i)\frac{\partial^2\ln h_s(t;\bm{\theta})}{\partial\theta_k \, \partial\theta_\ell}\biggr|_{\hat{\bm \theta}}, \\
H'_{k\ell} =& \sum_i \hat w_s^2(m_i) \left(\frac{\partial\ln h_s(t;\bm{\theta})}{\partial\theta_k} \frac{\partial\ln h_s(t;\bm{\theta})}{\partial\theta_\ell}\right)\biggr|_{\hat{\bm \theta}}, \\
E_{k(xy)} =& \sum_i \frac{\partial w_s(m_i)}{\partial W_{xy}}\biggr|_{\widehat W_{ss},\widehat W_{sb},\widehat W_{bb}} \frac{\partial\ln h_s(t;\bm{\theta})}{\partial\theta_k}\biggr|_{\hat{\bm \theta}}, \\
C'_{(xy)(uv)} =& \sum_i \frac{g_x(m_i) \, g_y(m_i) \, g_u(m_i) \, g_v(m_i)}{\big(\hat N_s \, g_s(m_i) + \hat N_b \, g_b(m_i)\big)^4},
\end{align*}
where $(xy)$ and $(uv)$ iterate over $\{ ss, sb, bb \}$, and $\hat w_s(m_i) = w_s(m_i; \hat W_{ss}, \hat W_{sb}, \hat W_{bb})$. The asymptotically correct expression for the binned approach is also derived in Ref.~\cite{Langenbruch:2019nwe}.

The first term of \eq{cov_corr} is the covariance for a weighted score function as described by \eq{weighted_score} with independent weights $w_i$. The second term is specific to \sweights and always reduces the covariance of the estimate $\hat \theta$. This reduction is caused by the fact that the \sweights are estimated from the same data sample. If the shapes of $g_s(m)$ and $g_b(m)$ are also estimated from the data sample, \eq{cov_corr} has to be extended with further terms, see Appendix~\ref{app:sampleestimates}.

\section{Practical applications of COWs and sWeights}\label{sec:application}

All of the studies in this section are available to view online at Ref.~\cite{sweights_implementation}. This includes generic implementations of extracting \emph{sWeights} (Sec.~\ref{sec:sweights}) and COWs (Sec.~\ref{sec:cows}) with the variants detailed in this
document, as well as a class which performs a correction to the covariance matrix when fitting unbinned weighted data (Sec.~\ref{sec:uncerts}). The interface is provided in python and offers
support for probability distribution functions defined
in either \texttt{scipy}~\cite{scipy}, \texttt{ROOT} (via \texttt{TTrees})~\cite{root} or \texttt{RooFit}~\cite{roofit}. We also point out that the \texttt{RooStats}~\cite{roostats} package implements what we here call \sweights Variant B but does not implement the other variants or COWs.

An important point to remember is that the derivation of the \textit{sWeight} formalism in Sec.~\ref{sec:sweights} simply requires a sensible estimate for the signal and background shapes, $\hat{g}(m)$. It does not require any special refitting or yield-only fitting which has been commonly recommended in other \textit{sWeights}
discussions. Using the formalism outlined in this article, one only needs to fit the discriminant variable(s) (usually
a candidate invariant mass) once; with the freedom to float, fix or constrain any parts of the shape or yields therein to obtain
$\hat{g}(m)$. One can then extract the \sweights for any component of $\hat{g}(m)$ and need not be concerned about fixed or constrained
yield parameters. Moreover, the range used to compute the weights can even be different from the one used to extract the weights and indeed one could even use a binned fit (\eg if the sample is large) to obtain estimates of the \pdfs and still extract per-event \sweights.
This formalism also allows one to extract the pure weight function, i.e.~one that is valid for any value of the mass not just a weight
per event.
In the case of extracting COWs a fit never even needs to be performed, one simply needs estimates for $\hat{g}_s(m)$, $\hat{g}_b(m)$ and $I(m)$. As described in Sec.~\ref{sec:cows_in_wild} these can be obtained from the data sample directly for $\hat{g}_s(m)$ and $I(m)$, and as a sum of polynomials for $\hat{g}_b(m)$.
As we
will see in the practical examples below there are some pitfalls to be wary of and we would always recommend that each use case follows
a similar approach to that shown here: produce ensembles of simulated events to check that biases are small and variances are
as expected.

\subsection{Statistical test of independence}

An important prerequisite for the extraction of \sweights is that the data samples for the discriminant and control variables are statistically independent for both signal and background; which means that the total \pdf factorises for the discriminant and control variables. If this is not the case then the extracted \sweights can be biased. The COWs formalism, described in Sec.~\ref{sec:cows}, allows one to overcome this by expanding the \pdf into a series of terms which do factorise. In order to check the independence in a data sample we recommend use of the Kendall rank correlation coefficient~\cite{kendall:1975}. A simple function to compute the correlation coefficient, $\tau$, is provided in Ref.~\cite{sweights_implementation}. It should be noted that the uncertainty on $\tau$ scales approximately with $1/\sqrt{N}$, where $N$ is the sample size.

\subsection{A simple example comparing \textit{sWeight} variants}
\label{sec:app:simple}

\begin{figure}
    \centering
    \includegraphics[width=0.48\textwidth]{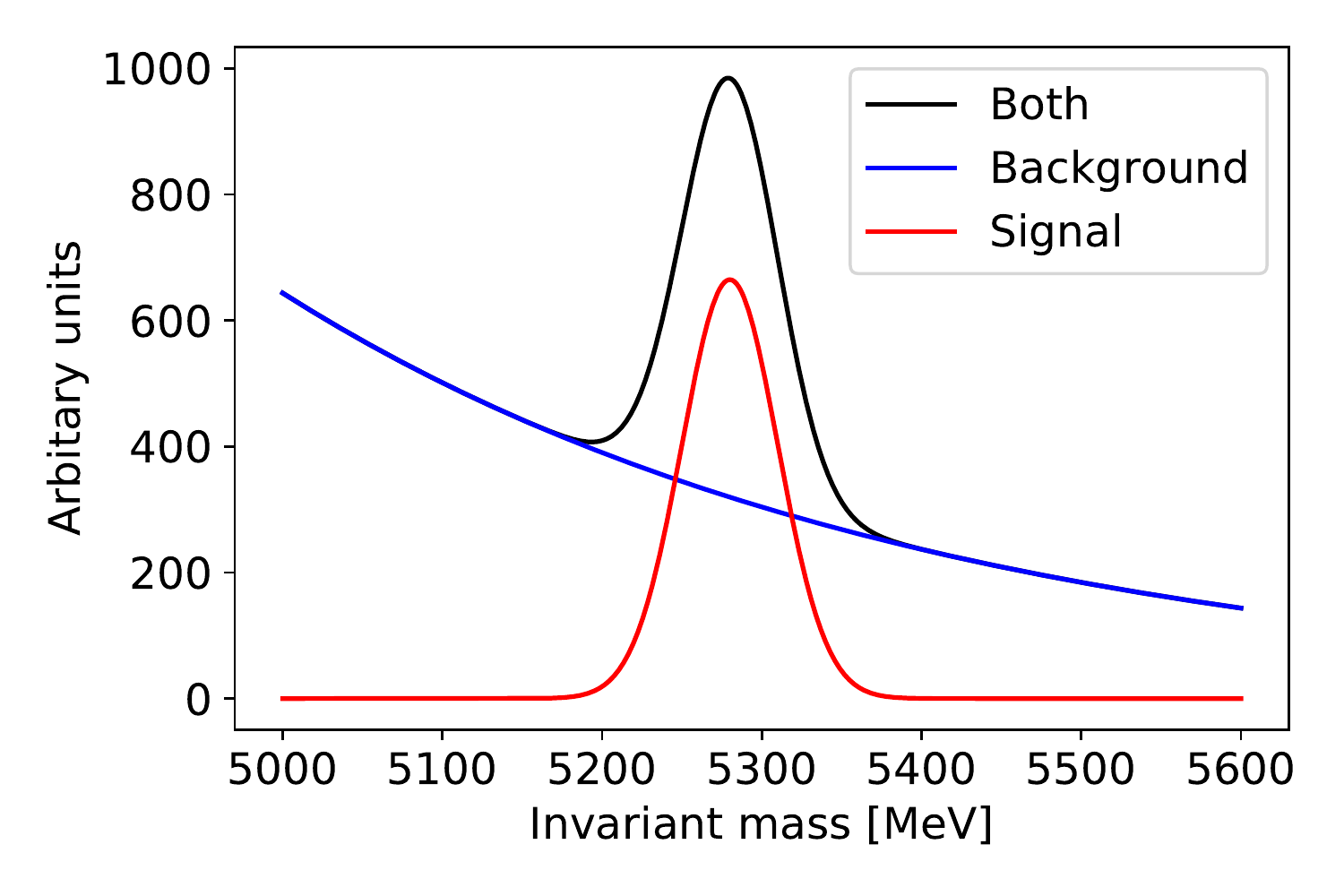} \\
    \includegraphics[width=0.48\textwidth]{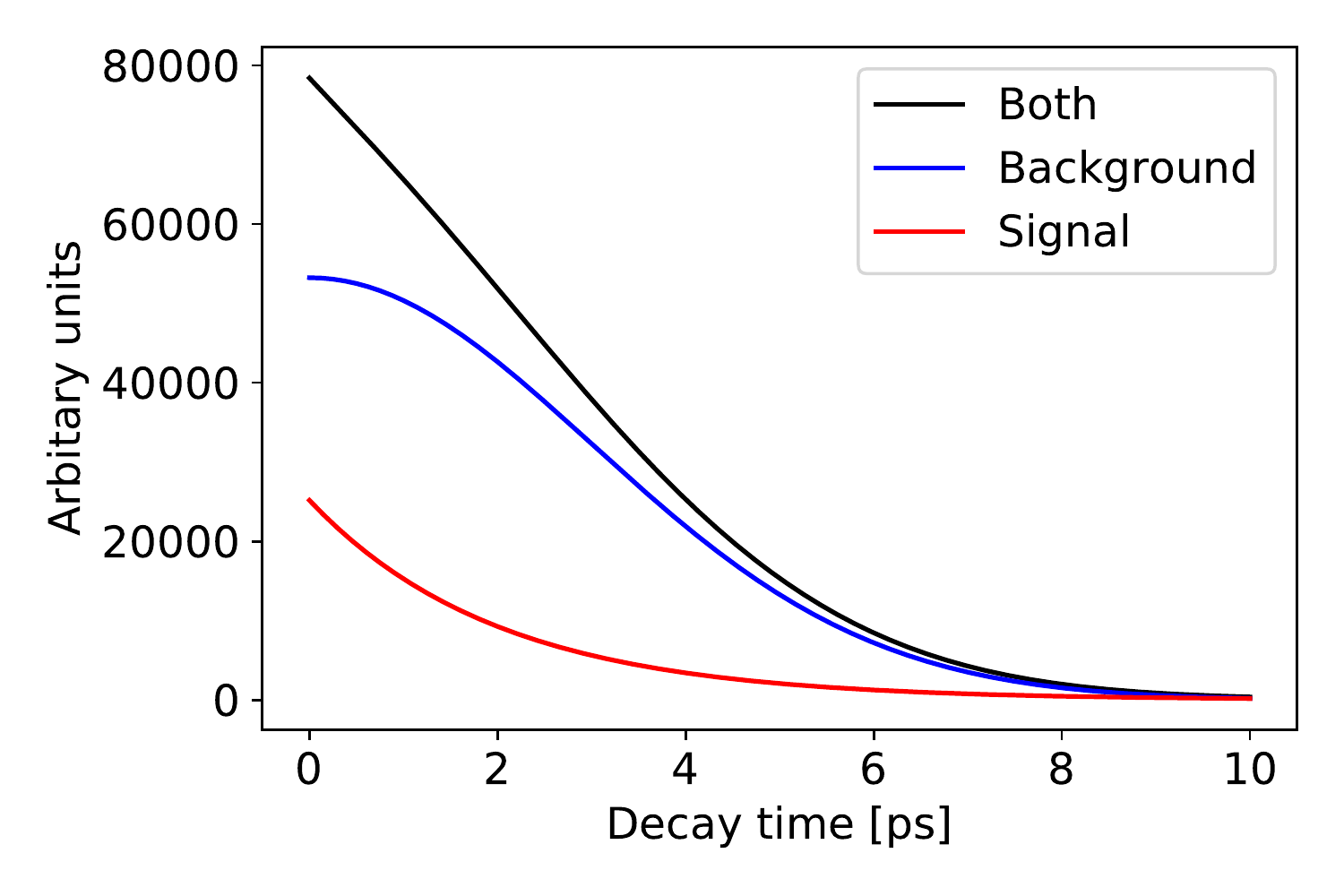}
    \caption{The $m$ (top) and $t$ (bottom) projections of the \emph{true} distributions used to generate the pseudo-experiments studied in Sec.~\ref{sec:app:simple}.}
    \label{fig:ex1_truth}
\end{figure}

A simple example has been considered to demonstrate the method and illuminate some of the
small differences between the variants described in
Sec.~\ref{sec:app_fin_samps}. A common application of \sweights in flavour physics is to extract the lifetime
of a candidate using its invariant mass to isolate it from the background.
In this example we take two independent variables; invariant mass $m$ and decay time $t$
of a $B$-meson candidate. Our observed dataset contains an arbitrary mixture of signal; normally (exponentially) distributed in $m$ ($t$), and background; exponentially (normally) distributed in $t$ ($m$), events. The $m$ and $t$ projections of the \pdf, which is the $f(m,t)$ of \eq{fmt}, used to generate simulated events is shown in
\fig{ex1_truth}.

For each simulated dataset, the estimates $\hat{g}_s(m)$ and $\hat{g}_b(m)$ are obtained by fitting back the generated mass distribution.
 We then compute the
$\widehat W_{xy}$ matrices of Eqs.~\ref{eq:wxya},~\ref{eq:wxyb} and~\ref{eq:wxyc} for variants A, B and C respectively. Finally, the weight functions, both $\hat{w}_s(m)$ and $\hat{w}_b(m)$, are extracted for each variant using \eq{hat_w_s}.
Within variant C
we extract the weight functions using both of the methods described in Sec.~\ref{sec:maxlh}: i) by twice inverting the covariance matrix and ii) by using Eq.~\ref{eq:dir_cov} on the covariance of a fit in which only the yields float.

\begin{figure}
    \centering
    \includegraphics[width=0.48\textwidth]{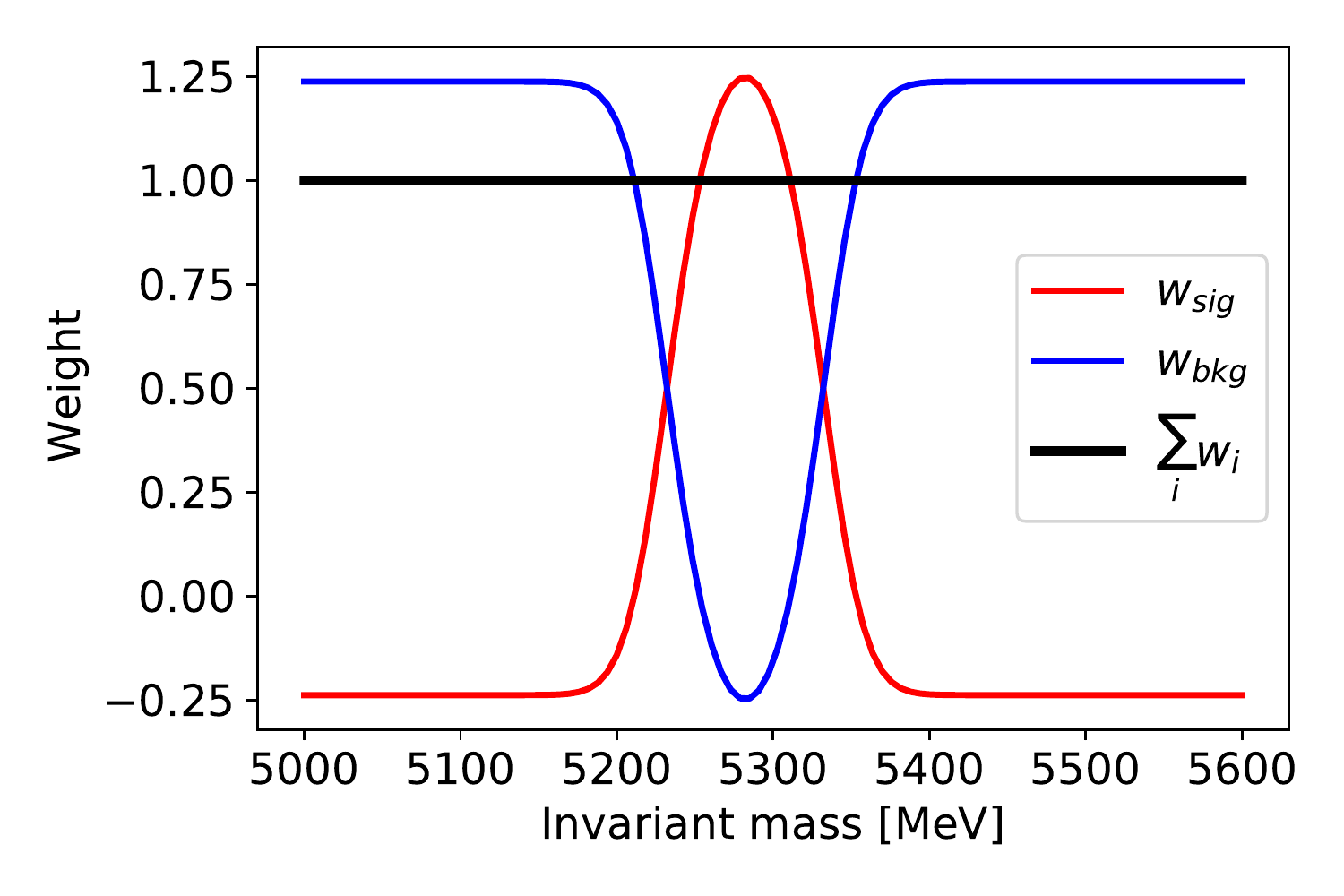}
    \caption{Distribution of the weight functions, $w_s(m)$ (red) and $w_b(m)$ (blue), as well their sum (black), extracted using the Variant B method. The other variants give very similar looking results. }
    \label{fig:ex1_weights}
\end{figure}

\begin{figure}
    \centering
    \includegraphics[width=0.48\textwidth]{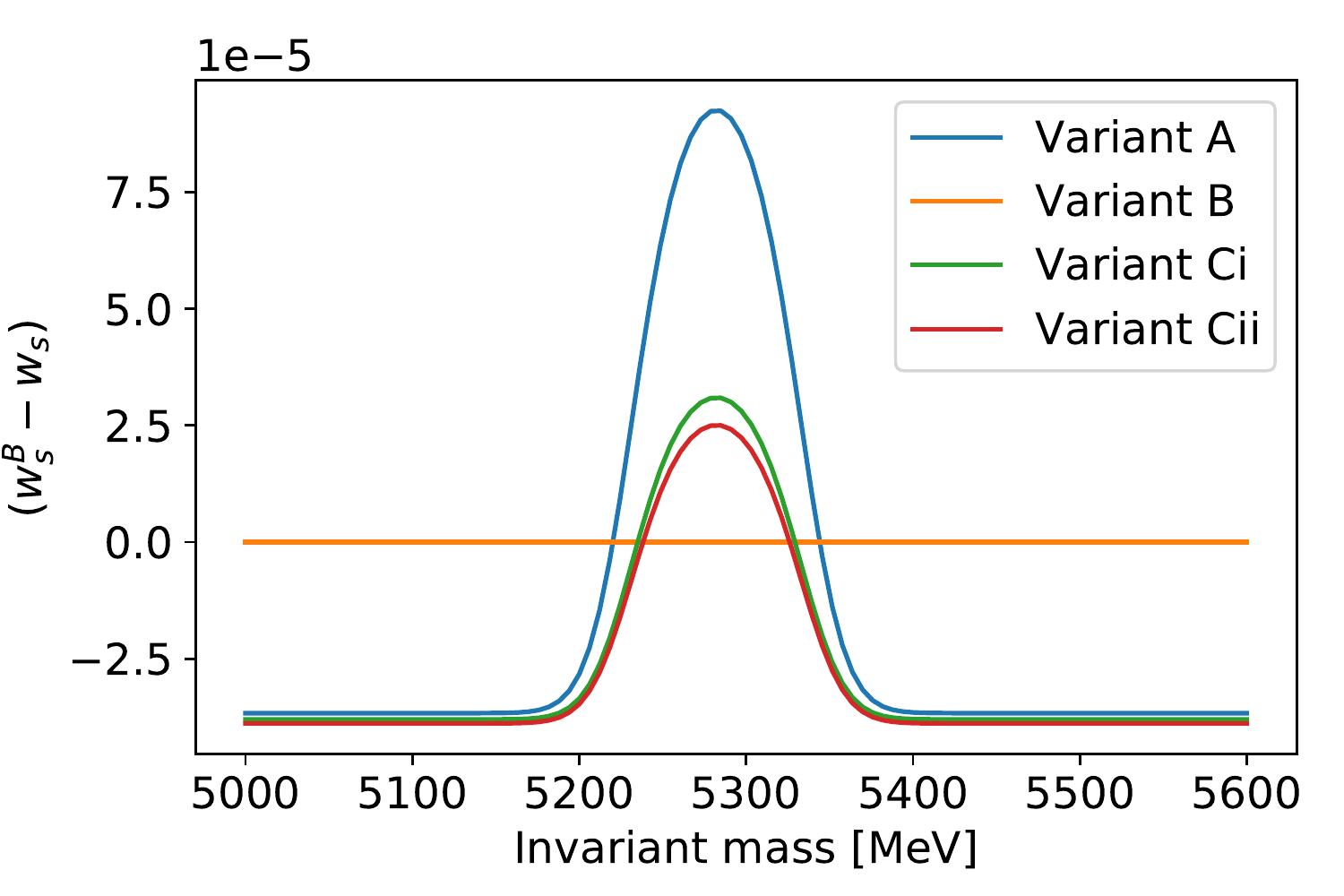}
    \caption{Difference between the extracted signal weights, $w_s(m)$, from each variant with Variant B as the reference.}
    \label{fig:ex1_weightdiffs}
\end{figure}

The distribution of the weight functions, $w_s(m)$ and $w_b(m)$, as a function of the discriminant variable, invariant mass, are shown for the nominal Variant B method in \fig{ex1_weights} for one pseudo-experiment containing 5K (20K) signal (background) events. The other variants give very similar looking distributions, although small differences can be seen when inspecting their relative differences as shown in \fig{ex1_weightdiffs}. It is useful to confirm the formalism of Sec.~\ref{sec:sweights} with a numerical evaluation of this
example. Indeed we see, with all four of the methods inspected here, that $\int w_i(m) g_j(m) dm = \delta_{ij}$, as well as $\sum_i w_i(m)=1$ for all $m$, to a high numerical precision. We also evaluate the sum of weights and sum of squared weights in order to make a comparison with
the yield estimates and uncertainties extracted from the discriminant variable fit (for a proof that the sum of squared weights provides an estimate for the asymptotic variance see Appendix~\ref{sec:exp_var_sum_w}). The results are shown in Table~\ref{tab:ex1_weights}, along with those from the free fit and a fit with only the yields floating. This
demonstrates Eq.~\ref{eq:sum_w_s_equal_nz} for Variant B, \ie that the fitted yield is exactly reproduced by the sum of weights. Whilst at first glance the sum of squared weights may appear to underestimate the variance of the fitted yield, one has to realise that the weights are agnostic of any variance in the shape parameters. Table~\ref{tab:ex1_weights} shows that the
sum of squared weights accurately reproduces the variance of a fit in which only the yields float.

\begin{table}
    \centering
    \renewcommand{\arraystretch}{1.3}
    \begin{tabular}{lcccc}
    \hline
    \hline
    \textbf{Fit methods} & $N_s$ & $\sigma(N_s)$ & $N_b$ & $\sigma(N_b)$ \\
    \hline
    EML Fit (all pars.)               & 49591.22 &   351.23 & 200409.16 &   523.61 \\
    EML Fit (yields only) & 49591.22 &   311.25 & 200409.16 &   497.69 \\
    \hline
    \hline
    \textbf{sWeight methods}     &   $\sum w_s$ &  $\sqrt{\sum w_s^2}$ & $\sum w_b$ & $\sqrt{\sum w_b^2}$ \\
    \hline
    Variant A                & 49591.01 &   311.26 & 200408.99 &   497.70 \\
    Variant B                & 49591.22 &   311.25 & 200409.16 &   497.69 \\
    Variant C                & 49595.97 &   311.24 & 200408.98 &   497.67 \\
    Variant D                & 49596.17 &   311.24 & 200410.08 &   497.67 \\
    \end{tabular}
    \caption{A comparison of the fitted component yields and the errors from the fit with the extracted sum of weights and sum of weights squared. This numerically demonstrates why we recommend Variant B as the best choice as the weights precisely reproduce both the central value and uncertainty of the fitted yield. }
    \label{tab:ex1_weights}
\end{table}

\begin{figure}
    \centering
    \includegraphics[width=0.48\textwidth]{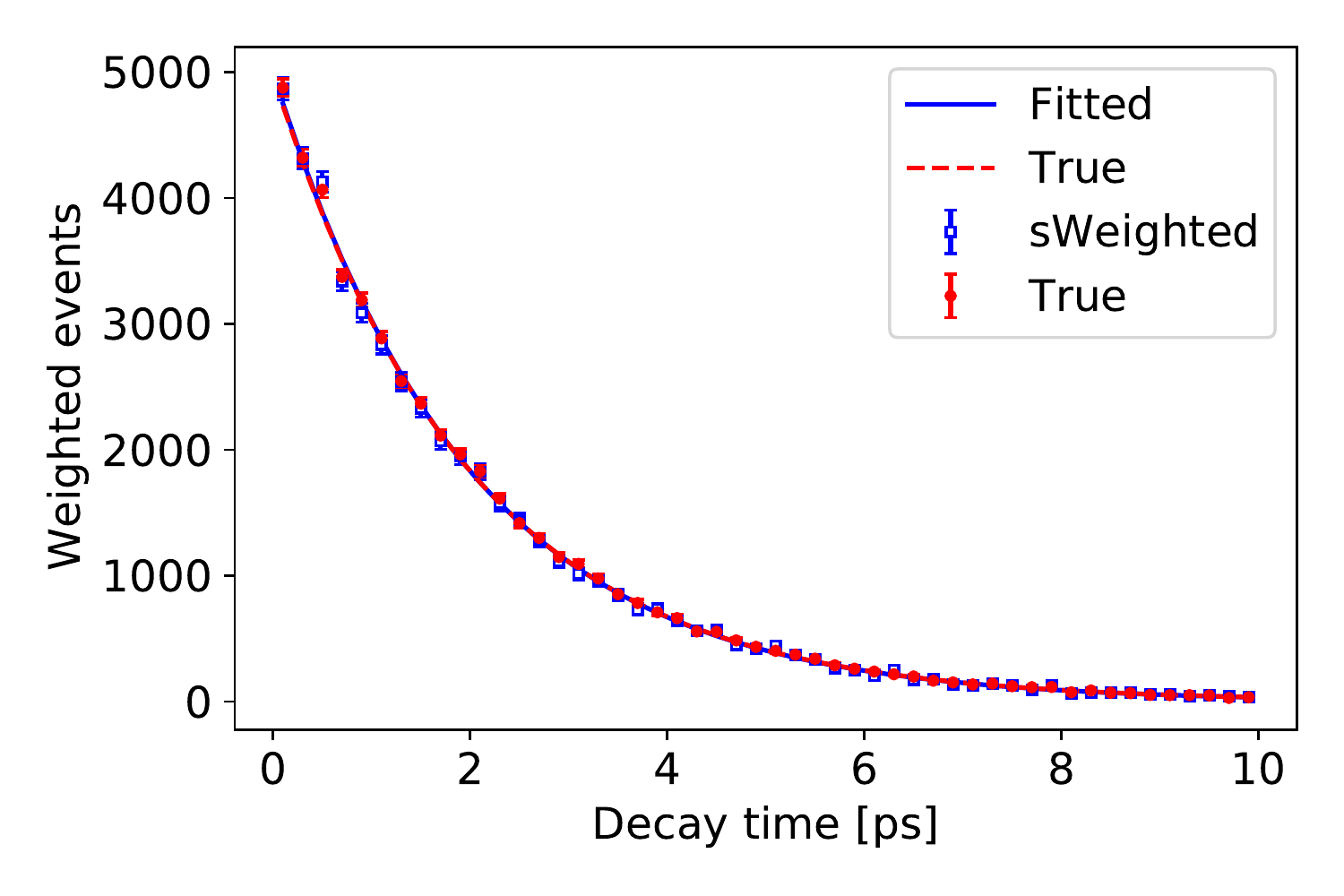}
    \caption{The decay-time distribution of true signal candidates (red points) and the total signal and background dataset, weighted with $w_s(m)$ (blue points), extracted using Variant B. The solid blue line shows the result of an exponential fit to the weighted distribution. The dashed red line shows the true underlying decay-time distribution used to generate the dataset from. Weights extracted using the other variants give very similar looking results. }
    \label{fig:ctrl_wts}
\end{figure}

Finally, we apply the signal weights to our dataset in the control dimension, $t$, and fit this with the expected exponential distribution. We subsequently find that we obtain an accurate estimate of the shape, $\hat{h}_{s}(t)$, finding that the slope parameter has a very similar value
to that which would have been obtained had we performed the fit in two dimensions to start with.
The weighted and true distributions in the control variable $t$ are shown in \fig{ctrl_wts} for Variant B. The other variants produce very similar looking distributions. The fitted values of the exponential slope to the ($s$)weighted data for each variant, compared to that obtained from a full 2D fit, are given in Table~\ref{tab:ctrl_fits}. Note that the uncertainties on these parameters are appropriately scaled
according to the description given in Sec.~\ref{sec:uncerts} as we are now fitting weighted data.

\begin{figure}
    \centering
    \includegraphics[width=0.48\textwidth]{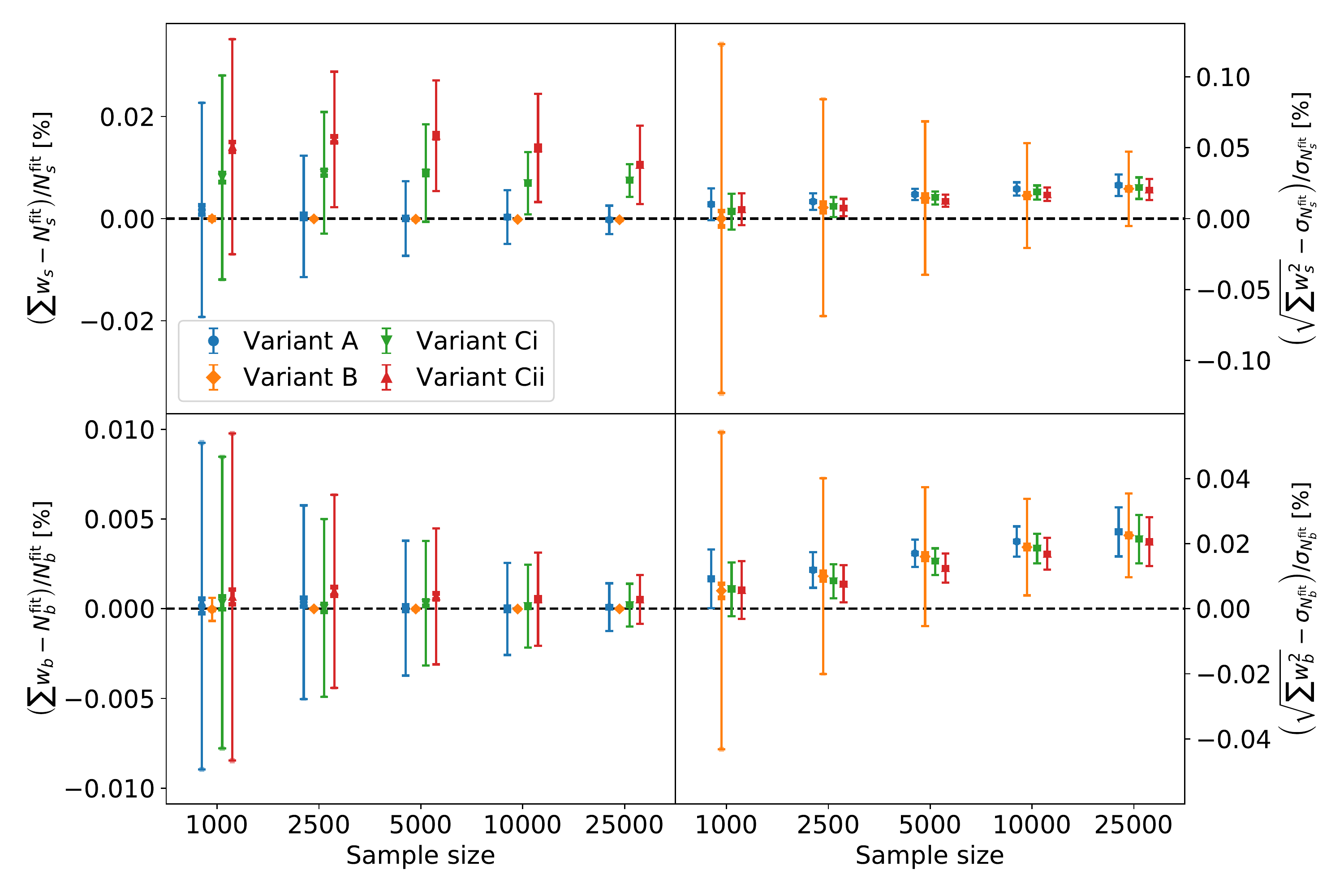}
    \caption{The upper (lower) left plots show the percentage difference between the sum of weights and the fitted yield, from a fit to the discriminant mass variable in which only the yields float, for the signal (background) components. The right plots show the percentage difference between the square root of the sum of squared weights and the error on the fitted yield. The points (with thin error bars) show the mean (width) of the distribution across the ensemble of pseudo-experiments. The thick error bars (shaded boxes) represent the standard error on the mean (width) across the ensemble.}
    \label{fig:ex1_toy_ylds}
\end{figure}

\begin{figure}
    \centering
    \includegraphics[width=0.48\textwidth]{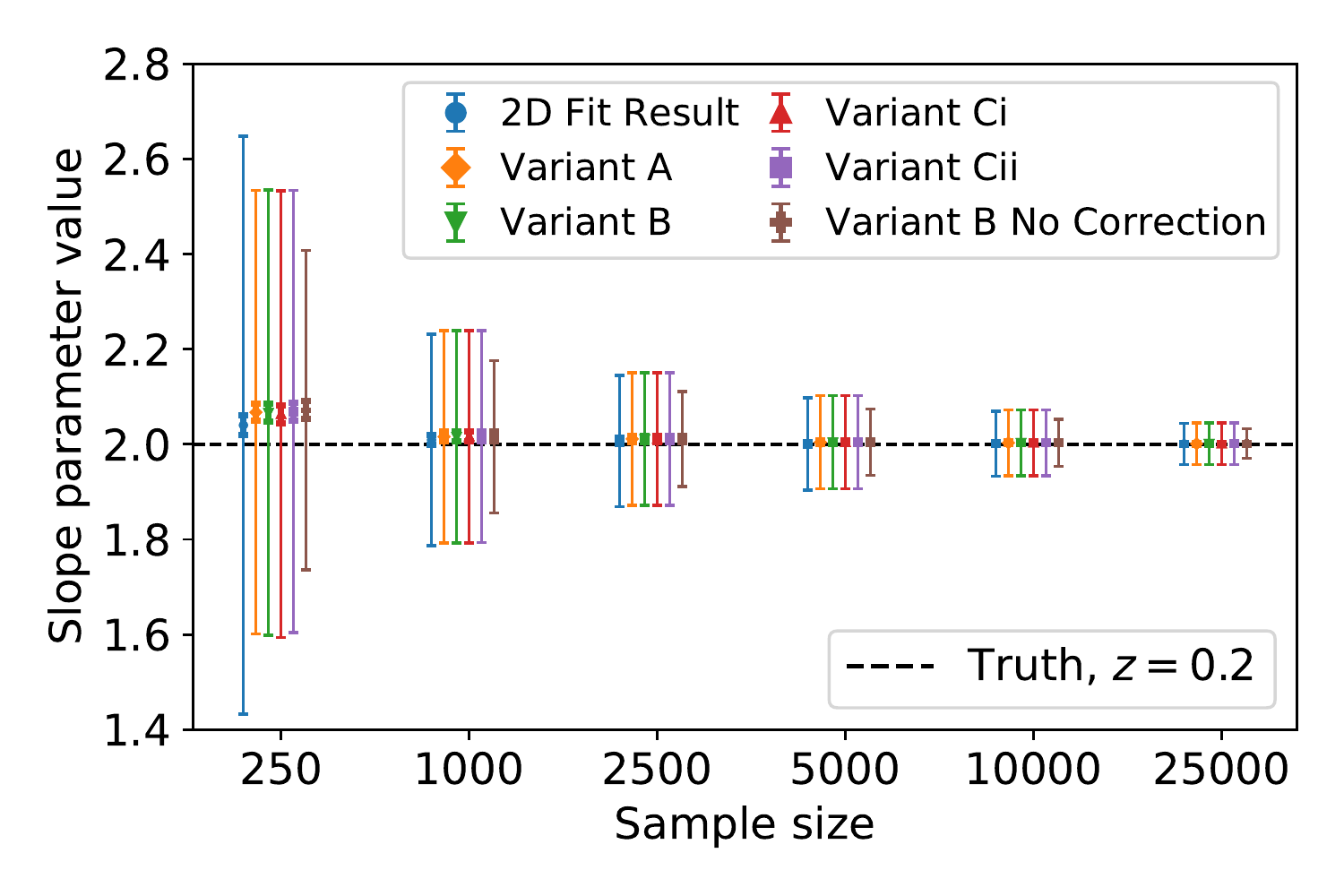} \\
    \includegraphics[width=0.48\textwidth]{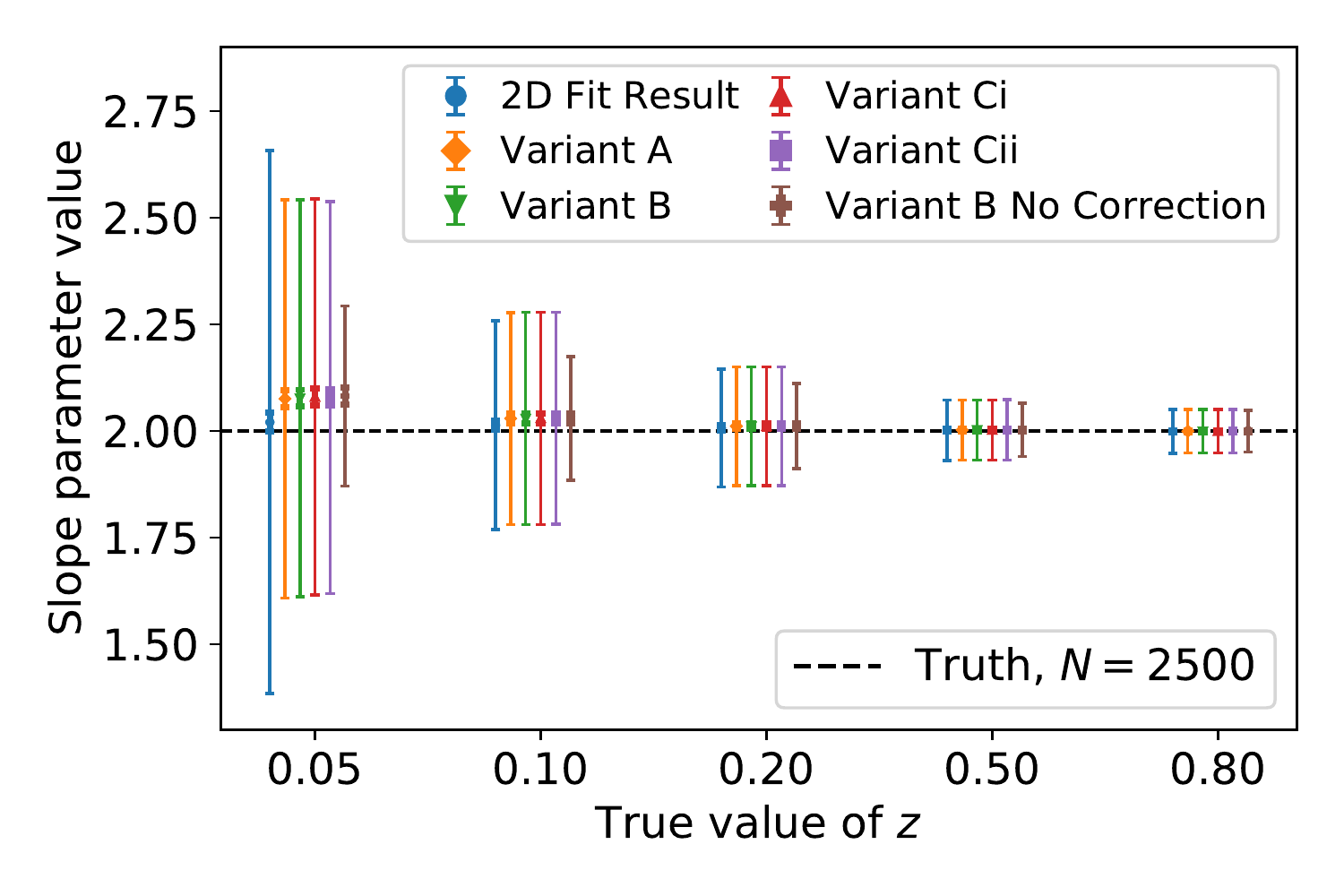}
    \caption{A comparison of the performance of each \emph{sWeight} variant with a full two-dimensional fit as a function of the sample size (top) and signal to background ratio, explicitly the value of $z$, (bottom). For the top figure $z=0.2$ and for the bottom figure the sample size is $N=2500$. The points (with thick error bars) show the mean (standard error on the mean) of the distribution of fitted slope values across the ensemble of pseudo-experiments. The thin error bars shows the square root of the mean of the variances of the fitted slope extracted across the ensemble. Note neither of the $x$-axes are on a linear scale. }
    \label{fig:ex1_toys}
\end{figure}

We then repeat this study on ensembles containing 500 pseudo-experiments in order to ensure that any of the behaviour seen is not just a fluke of the specific dataset shown in this example. We also perform the same study on ensembles with smaller samples sizes and with different signal to background ratios, the results are shown in Figs.~\ref{fig:ex1_toy_ylds} and~\ref{fig:ex1_toys}. We find that each of the variants described here give very similar results and can accurately reproduce the full two-dimensional fit with, at least in this case, a minimal loss in precision.

Figure~\ref{fig:ex1_toy_ylds} shows that the sum of weights (left two panels) for Variant B accurately reproduce the fitted yield. Variant A is also unbiased in this respect but has a slightly larger spread (note the very small y-axis), whilst Variants Ci and Cii give a very small bias and tend to overestimate the yield by about 0.1 per mill. When inspecting the variance properties, sum of squared weights (right two plots), we can see that all of the methods tend to very slightly over estimate the fit uncertainty. Variant B has a much larger spread of variances than the other methods which are all similar.

Figure~\ref{fig:ex1_toys} shows the importance of computing the covariance matrix correction using \eq{cov_corr} (a comparison of the brown points with the rest). For very small amounts of signal, either small overall sample size or small values of the signal to background ratio, we see some slight biases and a much smaller average uncertainty when using the weights method, as compared to the full two-dimensional fit. Inspection of the studentised residual distributions suggest a small amount ($\sim10\%$) of under-coverage in these cases, which is more than likely due to the asymptotic assumptions made when correcting the covariance matrix no longer being valid.

\begin{table}
    \centering
    \begin{tabular}{ c c  }
      Method    & Fit Result \\
      \hline
      2D Fit    & $2.0025 \pm 0.0137$ \\
      Variant A & $2.0067 \pm 0.0138$ \\
      Variant B & $2.0067 \pm 0.0138$ \\
      Variant C & $2.0068 \pm 0.0138$ \\
      Variant D & $2.0068 \pm 0.0138$

    \end{tabular}
    \caption{A comparison of the fitted values for the control variable slope when fitting to the sWeighted data sample with the outcome if a full 2D fit had performed. Comment on loss of precision. Note that the weighted fits do not have the Christoph correction included so are almost certainly wrong.}
    \label{tab:ctrl_fits}
\end{table}

\subsection{A more complex example with Variant B}

\begin{figure}
    \centering
    \includegraphics[width=0.48\textwidth]{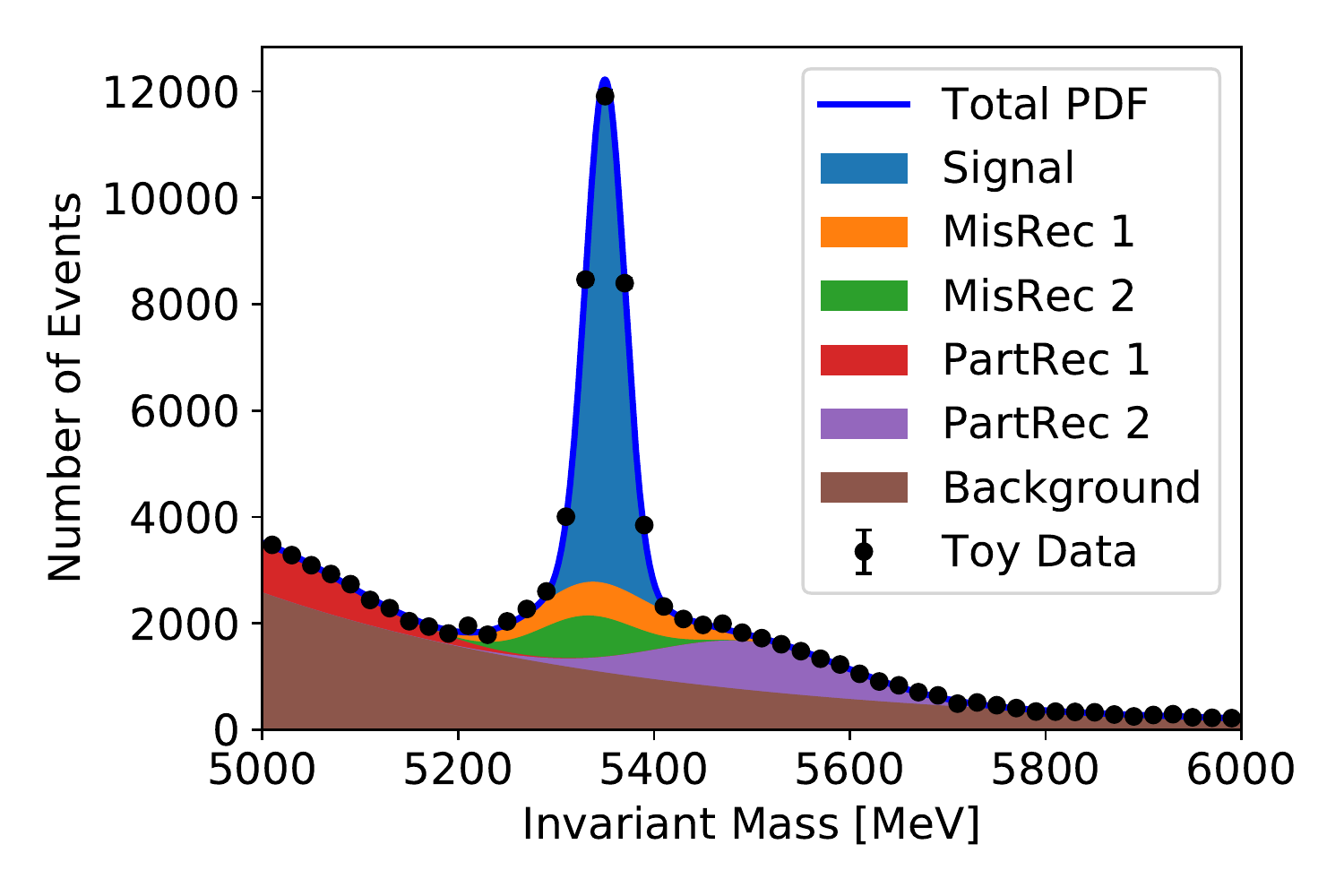}
    \caption{The probability distribution functions for the discriminant variable in the more complex example.}
    \label{fig:ex2_mdist}
\end{figure}

\begin{figure}
    \centering
    \includegraphics[width=0.48\textwidth]{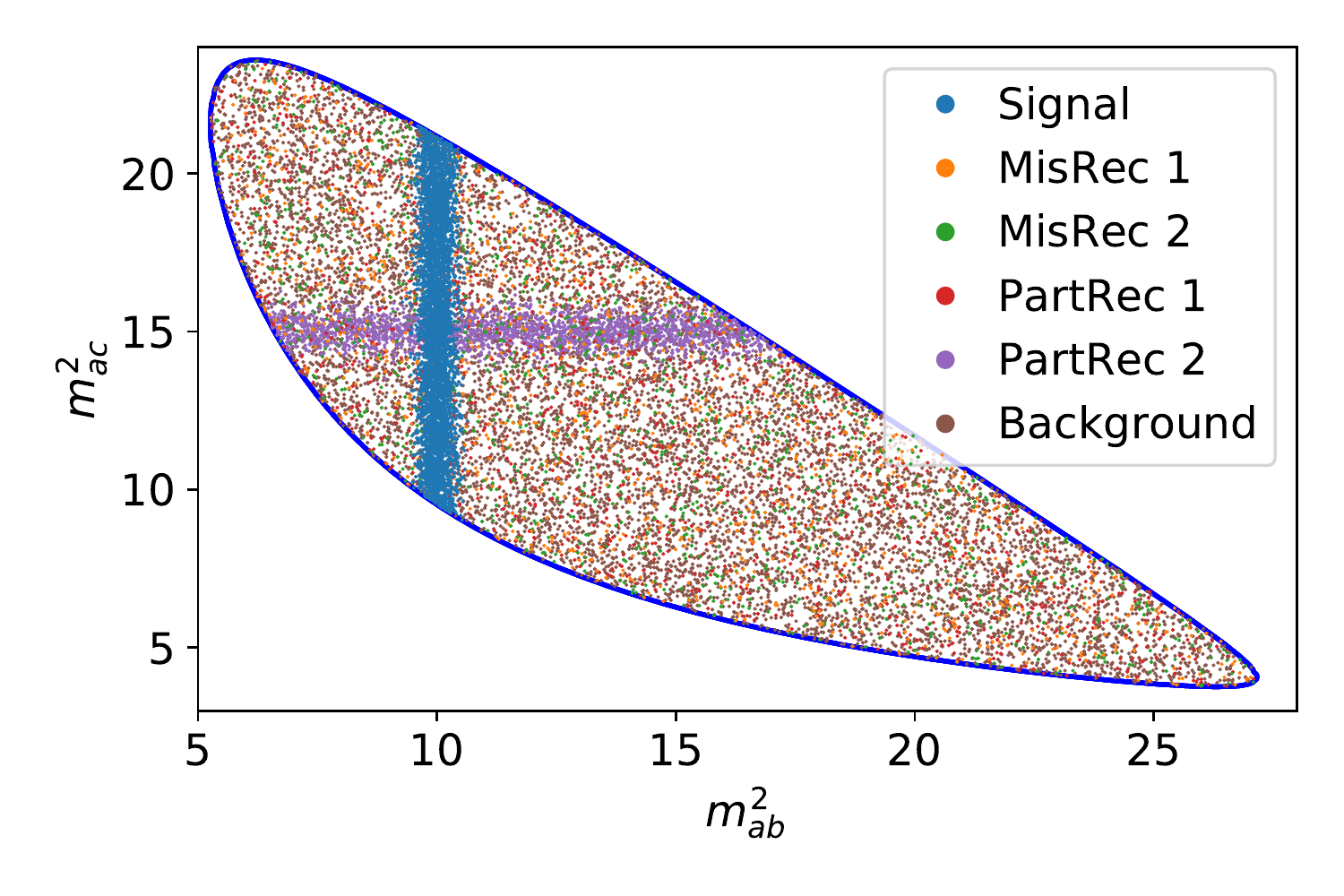} \\
    \caption{The true distributions of the control variables in the more complex example. The Dalitz variables are flat for all components apart from the signal (blue) and one of the backgrounds (purple) which appear as the vertical and horizontal bands in the Dalitz plot, respectively.}
    \label{fig:ex2_cdist}
\end{figure}

\begin{figure}
    \centering
    \includegraphics[width=0.48\textwidth]{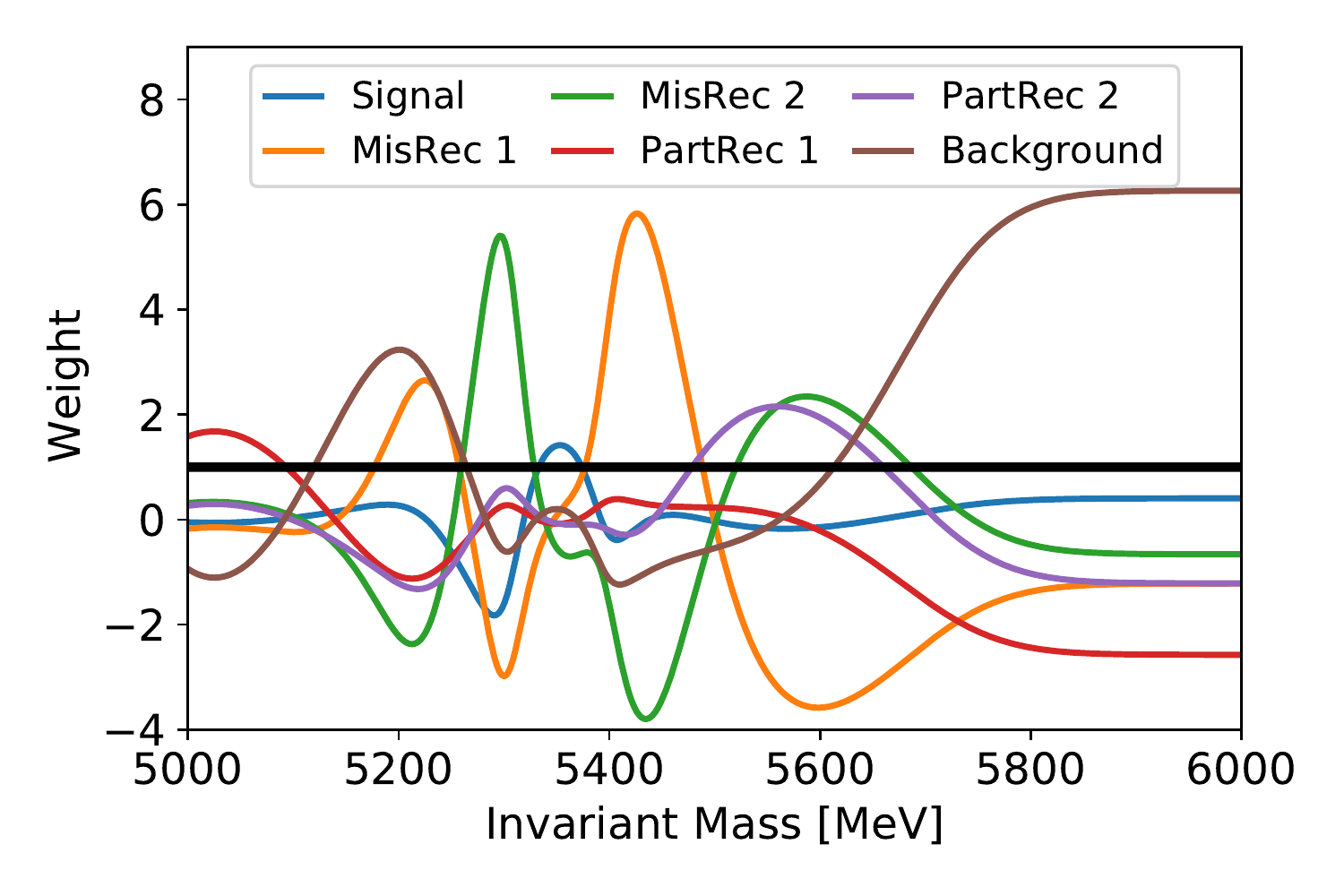}
    \caption{The distributions of the weight functions, $w_i(m)$, for
    each of the components in the invariant mass fit. Their sum is shown by the black line.}
    \label{fig:ex2_weights}
\end{figure}

In this section we test a more complex example for another common use case in flavour physics in which there are multiple different factorising components within $f(m,t)$ of which some may be signal and some may be backgrounds.
In this example we have an invariant
mass as the discriminant variable once more but now have six different components each with different \pdf{}s; some even peak under or near the signal in a similar way. For the control variable(s)
we use a simple discrete integer which labels the true component, $c\in[1,6]$, as well as two ``Dalitz" variables. We have assumed that
the discriminant invariant mass variable is constructed from a three-body decay of the form $X\to A B C$ and in this case the Dalitz variables are
the invariant mass squared of the $AB$ and $AC$ combinations.
We generate a pseudo-experiment from the true underlying model in which
the Dalitz variables are flat across the phase space for all components, apart from the signal which has a resonance in the $AB$ invariant mass, and one of the backgrounds which has a resonance in the $AC$ invariant mass, which appear as horizontal and vertical bands in the Dalitz plot. A visualisation of the generated dataset in the discriminant variable, $m$, is shown in Fig.~\ref{fig:ex2_mdist}. The control variable distributions are shown in Fig.~\ref{fig:ex2_cdist} where events have been coloured according to their true event type.
As in the previous example the generated dataset is fitted to obtain
estimates for $\hat{g}_i(m)$ and it is actually the result of this fit which is shown in Fig.~\ref{fig:ex2_mdist}. We then use the method of Variant B to obtain the $\widehat W_{xy}$ matrix (in this case a $6\times 6$ matrix), after which the 6 weight functions, $w_i(m)$, are extracted.
The distributions of these weight functions are shown in Fig.~\ref{fig:ex2_weights}.

\begin{figure}
    \centering
    \includegraphics[width=0.48\textwidth]{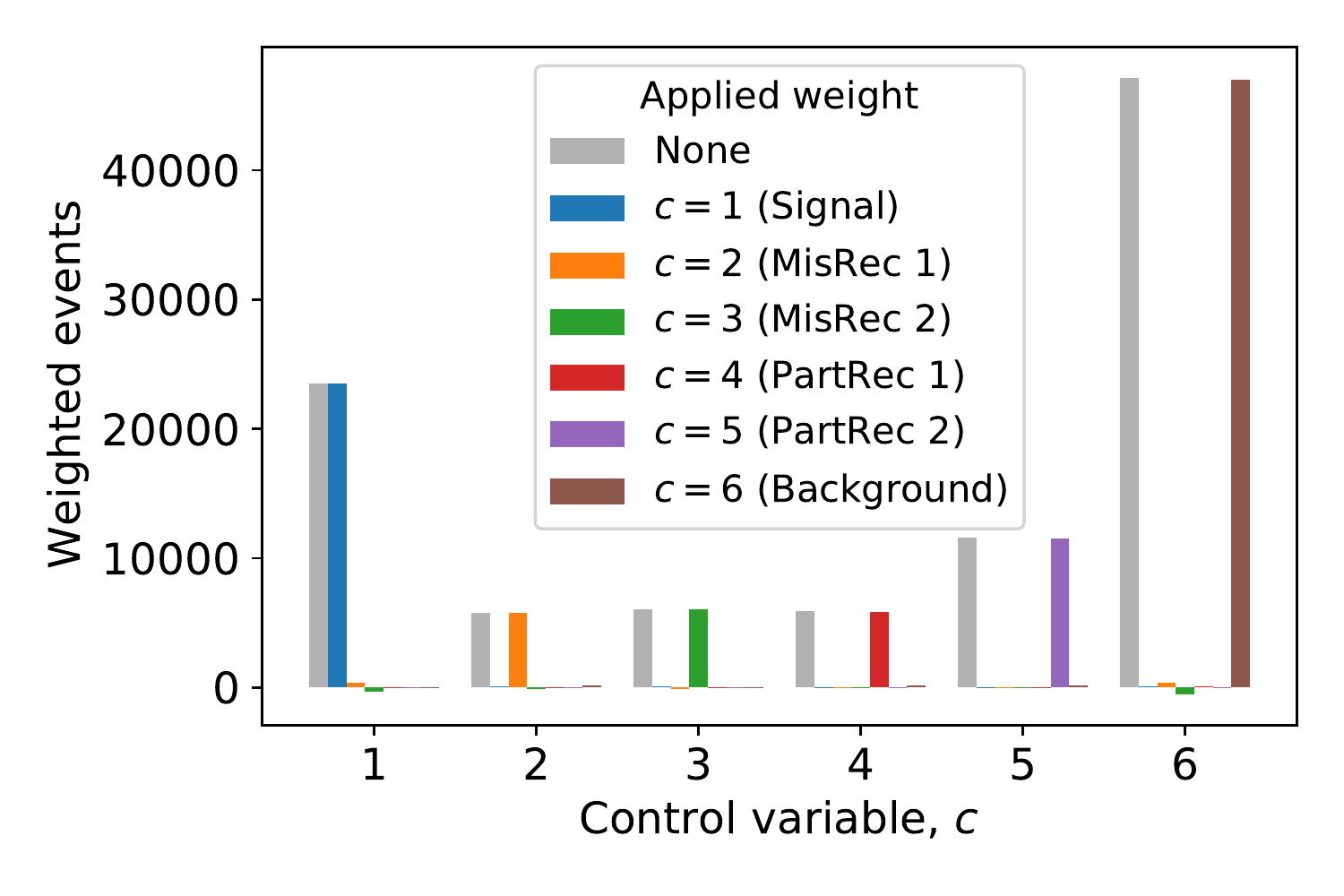}
    \caption{The weighted control variable, $c$, distribution before (grey) and after application of the relevant weight. The colour of each bar (bars have their $x$ position offset to aid the visualisation) represents the component weight that has been applied. One can see that the application of each separate weight accurately projects out the relevant component.}
    \label{fig:ex2_ctrl_wt}
\end{figure}

\begin{figure}
    \centering
    \includegraphics[width=0.48\textwidth]{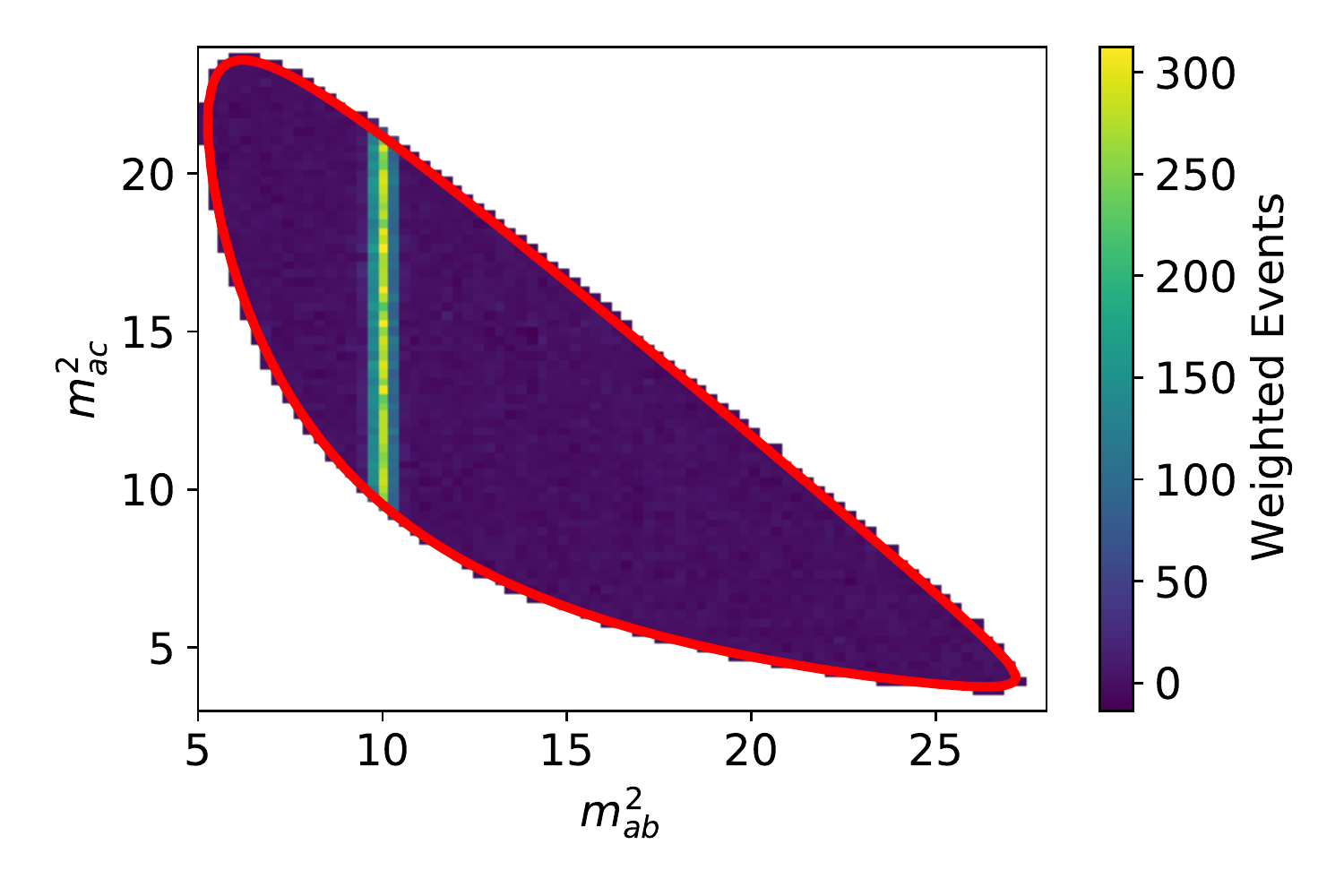}
    \caption{The weighted Dalitz distribution when applying the signal weight. One can see that the true distribution is very nicely recovered, with some fluctuations. Similar plots are found when applying the other components weights but are not shown.}
    \label{fig:ex2_ctrl_dlz}
\end{figure}

We can then inspect the distributions of the control variables when the various weights have been applied. One can see a very nice recovery of
the ``\emph{control}" variable in Fig.~\ref{fig:ex2_ctrl_wt} and the Dalitz variables for the signal component in Fig.~\ref{fig:ex2_ctrl_dlz}. The weighted Dalitz plots for the other components show a similar level of agreement with the truth. As seen before in Table~\ref{tab:ex1_weights} we again find in this example that $\int w_i(m) g_j(m) dm = \delta_{ij}$, $\sum_i w_i(m)=1$ for all $m$ and that the sum of weights and sum of squared weights accurately reproduce the corresponding fitted yield and variance.

This more complex example, in contrast to the previous simple case, exhibits rapidly oscillating weight functions (see Fig.~\ref{fig:ex2_weights}) which oscillate much more quickly than
the actual variation of the relevant component shapes themselves. This is because the weight is related to how the shapes overlap \emph{as well as} how they
vary themselves with mass. One can also see that competing (\ie similar) shapes oscillate out of phase, which is what we would expect as their yields are anti-correlated. It is worth noting that the sum of all component weights for any value of the discriminant variable, in this case invariant mass, is still unity.

It is worth highlighting that the components with the smallest yields have the largest amplitudes of the weight function. Clearly, this is because small yields will have large uncertainties and therefore will require a large variance of weights. This can then lead to fairly sizeable fluctuations in the weights for small contributing samples when inspecting a relatively fine grain phase space, like that of the Dalitz plot. When inspecting certain distributions it is possible to see artefacts of these fluctuations appearing as features in a distribution, for example a band might seem to appear in a Dalitz distribution when in reality it is just large fluctuations around zero. Clearly, minimising the size of these fluctuations is prudent as it is generally undesirable to have few events with large weights. However, this issue only arises when trying to project out control variables for components which have a very small yield in the discriminant variable. Therefore our recommendation is to proceed with caution if you are trying to use the sWeight method for a fit component which is considerably smaller than others in the fit.

\subsection{An example exploiting COWs with a non-factorising background and efficiency effects}

\begin{figure}
    \centering
    \includegraphics[width=0.48\textwidth]{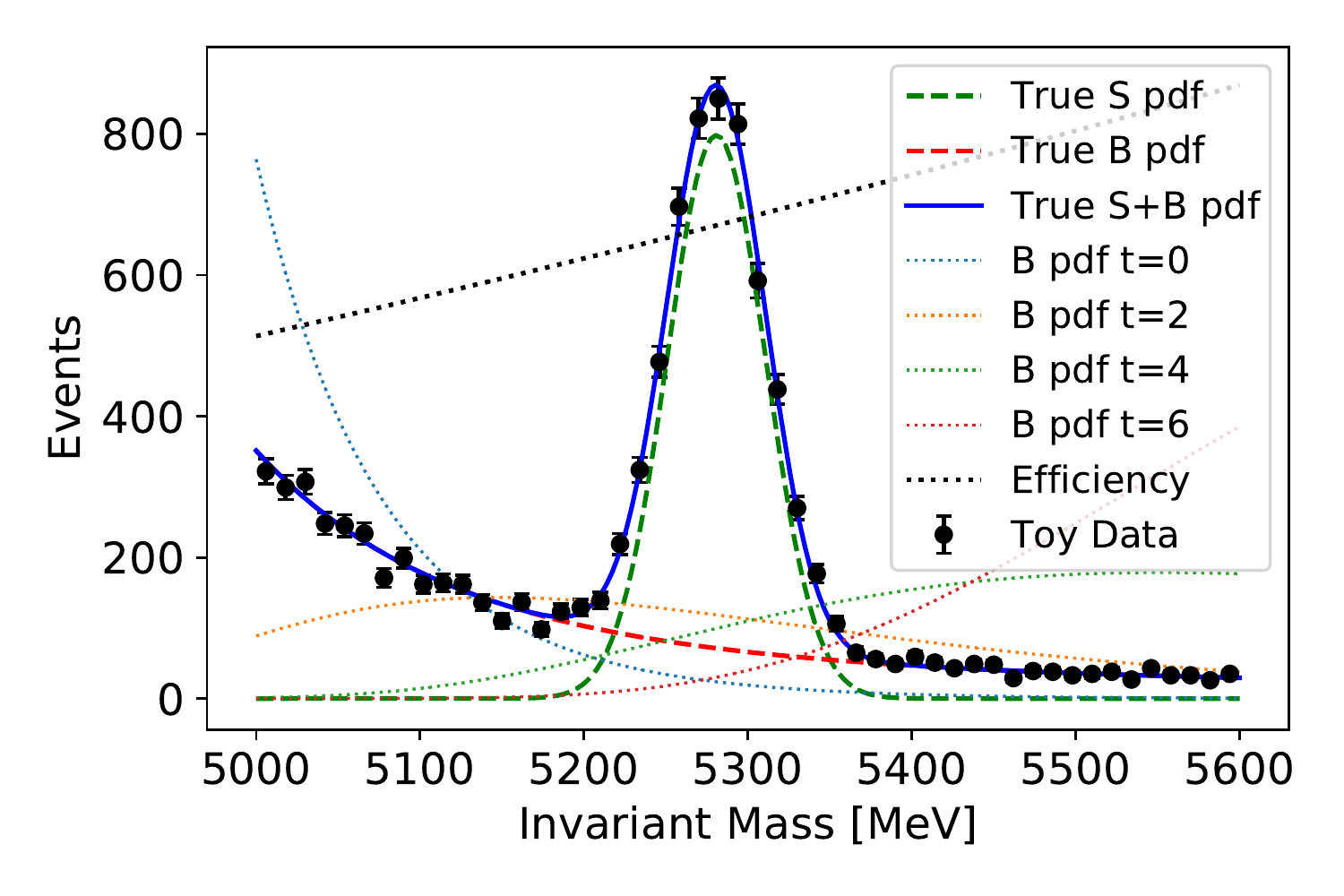} \\
    \includegraphics[width=0.48\textwidth]{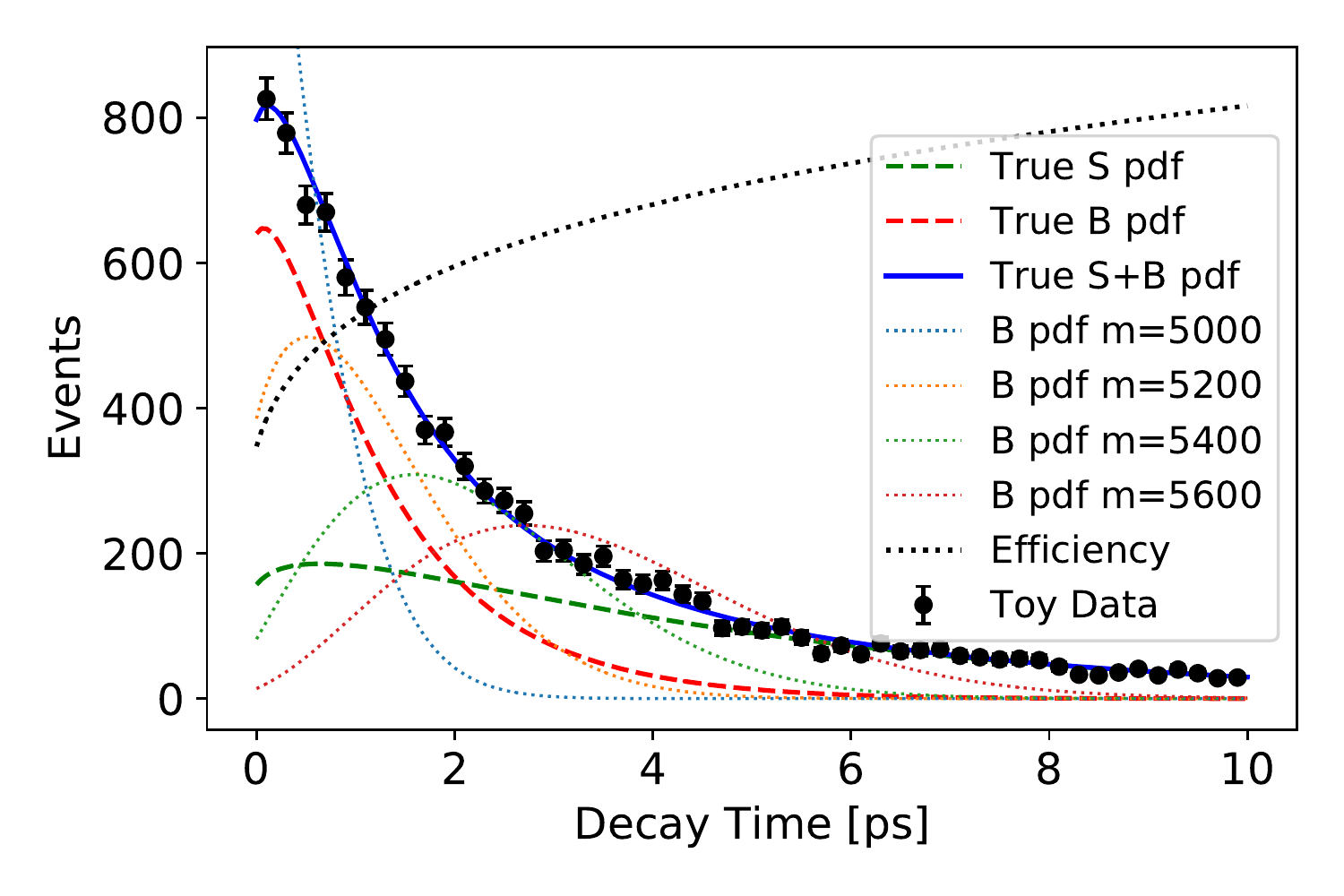}
    \caption{Projections of the true \pdf used in the extreme example case along with projections of the efficiency model.}
    \label{fig:ex4_eff}
\end{figure}

The final example we investigate considers an extreme case which has similar features to the first example (Sec.\ref{sec:app:simple}) but contains a highly non-factorising background model and a non-factorising efficiency. This emulates the use cases in which the signal efficiency is straightforward to estimate but the background efficiency is not.
The nature of the true model used to generate ensembles of experiments is shown in Fig.~\ref{fig:ex4_eff}, in which the non-factorising nature of the background is manifest in that the exponential slope of the background in mass varies with decay time, and both the mean and width of the normal distribution describing the background in decay time vary with mass. Projections of the integrated distributions along with the projection of the efficiency model used are also shown.

For this set of tests we perform an analysis on ensembles of simulated datasets using Variant B of the \sweights procedure described above along with various implementations of the COW formalism presented in Sec.~\ref{sec:cows}.
For the \sweights implementation the signal, $\hat{g}_s(m)$, and background, $\hat{g}_b(m)$ distributions are estimated by fitting the simulated sample as is done for the other examples above.
For the COWs implementation the same signal model, $\hat{g}_s(m)$, is used and a variety of tests are performed using:
\begin{itemize}
    \item The same estimate of the background as in the \sweights case, $\hat{g}_b(m)$
    \item Background functions given by sums of polynomials, up to 1st, 3rd and 5th order
    \item Variance functions of the COW equal to
    \begin{enumerate}
     \item unity, $I(m)=1$
     \item the true sum of \pdfs in mass as in \eq{cow_norm_func}, $I(m)=f(m) = zg_s(m) + (1-z)g_b(m)$, the COW equivalent of \sweights
     \item an estimation from the data sample itself using a histogram of the $1/\epsilon^2(m,t)$ weighted $m$-distribution as in \eq{cow_estimate}, $I(m)=q_B(m)$, where $B$ is the number of bins in the histogram, and binnings of 10, 25, 50 and 100 are tried.
    \end{enumerate}
\end{itemize}

\begin{figure}
    \centering
    \includegraphics[width=0.5\textwidth]{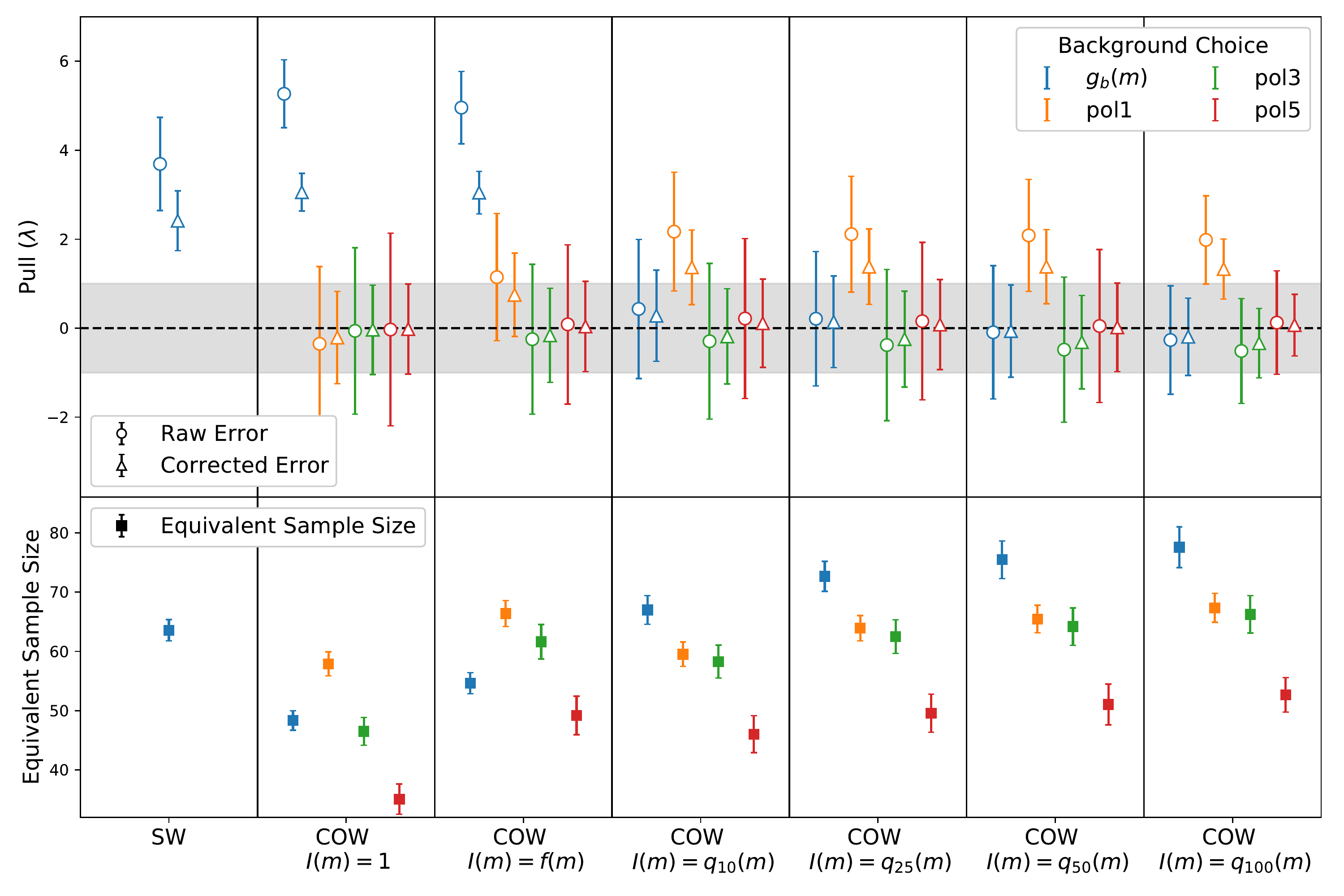}
    \caption{Results of the study incorporating a non-factorising efficiency model and a non-factorising background. The six panels, from left-to-right, show the equivalent sample size (bottom panels) and pull (top panels) of the fitted lifetime parameter on the weighted sample. The hollow triangles (circles) show the pull with (without) the covariance correction applied. The different colours represent different choices for the modelling of the background function, $g_b(m)$.  }
    \label{fig:ex4_res}
\end{figure}

The results for this analysis are shown in Fig.~\ref{fig:ex4_res} in which the simulated sample size is 2K events, with equal amounts of signal and background. We have also tested cases with different signal-to-background ratios and with different sample sizes and the conclusions are rather similar, apart from that fewer orders of polynomial are required to achieve a minimal bias when the sample size is smaller. It is also worth noting that for small samples ($< 100$ events) there are small biases due to the fact that the covariance correction of Sec~\ref{sec:uncerts} is only asymptotically valid.

It can be seen from Fig.~\ref{fig:ex4_res} that in the case of a highly non-factorising background model the traditional \sweights method can have a severe bias (first panel of Fig.~\ref{fig:ex4_res}). This is also the case for the COW formalism when $I(m)=f(m)$ or $I(m)=1$ (second and third panels of Fig.~\ref{fig:ex4_res}), neither of which contain the appropriate efficiency correction. This is overcome when using sums of polynomials which can effectively mitigate the non-factorising efficiency and non-factorising background. One can see that higher orders of polynomial achieve a smaller bias but reduce the statistical power of the method (the bottom panel of Fig.~\ref{fig:ex4_res} shows the equivalent sample size from the sum of signal weights with respect to the generated number of signal candidates).
When using $I(m)=q_B(m)$ these biases are significantly reduced, because in this case the estimate of $q_B(m)$ is suitably efficiency corrected. Small biases remain in this case if the background description is not sufficient (\eg in this case a first order polynomial is not enough).

Figure~\ref{fig:ex4_res} shows that when using the polynomial expansions for the non-factorising backgrounds the COWs formalism performs well, even in this extreme case, depending on the order of polynomial used in the background modelling and the form of the $I(m)$ variance function. With suitable choices of these, the bias can be minimised, with a price to pay in statistical precision (the higher order polynomial used the worse the precision, the fewer bins used and the smaller the sample used for the $q_B(m)$ estimate the worse the precision). It is clear that this choice will be analysis specific and it should be carefully considered on an individual basis. There will be a trade-off between systematic bias and statistical precision.

\section{Conclusions}

In summary this article gives a fresh overview and review of the \sweights method before discussing a generalisation of them which we dub ``Custom Orthogonal Weight functions" (COWs). We demonstrate that COWs can handle a variety of different applications and achieve statistically robust results with minimal loss in precision. Indeed COWs are applicable to situations in which the specific case of \sweights do not work.
\section{Acknowledgements}

The authors wish to thank their colleagues in the LHCb experiment, and members of the wider community of particle physicists with an interest in statistics, for the fruitful and enjoyable discussions which inspired this study. In particular they would like to thank Roger Barlow, Vladimir Gligorov and Louis Lyons. MK is supported by the Science and Technology Facilities Council (STFC), UK, under grant \texttt{\#ST/R004536/2}. HD acknowledges funding from the Deutsche Forschungsgemeinschaft (DFG – German Research Foundation) under award DE 3061/1-1. CL gratefully acknowledges support by the Emmy Noether programme of the Deutsche Forschungsgemeinschaft (DFG), grant identifier LA 3937/1-1. The authors would like to thank UK Research and Innovation (UKRI) for providing funds to allow open access.

\appendix
\section{Constrained minimization problem}\label{sec:constrained_opt}

We use Lagrange multipliers to find the function $w_s(m)$ which minimizes \eq{var_w_s} under the constraints \eq{normal} and \eq{orthogonal}. We need to find the extremum of
\begin{multline}
L(w_s(m), \alpha_s, \alpha_b) = \int w_s(m)^2 \, g(m) \de{m} - z^2 \\
- 2 \alpha_s \left(\intd m \, w_s(m) \, g_s(m) - 1\right) \\
- 2 \alpha_b \intd m \, w_s(m) \, g_b(m).
\end{multline}
The Lagrange multipliers $\alpha_{s,b}$ in $L$ were scaled by a factor of two without loss of generality. Since $L$ is a functional of $w_s(m)$, we need to use variational calculus. With
\begin{align*}
\delta \intd m \, w_s(m) \, \phi(m) &= \intd m \,  \delta w_s(m) \, \phi(m) \\
\delta \intd m \, w_s(m)^2 \, \phi(m) &= \intd m \, 2 w_s(m) \, \delta w_s(m) \, \phi(m)
\end{align*}
the variational score function is
\begin{multline}
\delta L = 2 \intd m \, \delta w_s(m) \Big[ w_s(m) \, g(m) \\
- \alpha_s \, g_s(m) - \alpha_b \, g_b(m) \Big] \overset{!}{=} 0.
\end{multline}
According to the fundamental lemma of calculus of variations, the equation is satisfied for any continuous $\delta w_s(m)$ only if the integrand inside the square brackets is zero. So we obtain
\begin{equation}
w_s(m) = \frac{\alpha_s \, g_s(m) + \alpha_b \, g_b(m)}{g(m)}.
\end{equation}

\section{Proof of self-consistency of \sweights calculated with variant B}\label{sec:sum_w_s_equal_nz}

Here, we prove \eq{sum_w_s_equal_nz}. For a more compact presentation, we use the definitions $s_i := \hat g_s(m_i)$, $b_i := \hat g_b(m_i)$, and $g_i := \hat g(m_i)$. We note that the hats are missing, but $s_i$, $b_i$, and $\mu_i$ are still computed from estimated \pdf{}s.

We insert \eq{hat_w_s} in the left-hand side of \eq{sum_w_s_equal_nz} and get
\begin{gather}
\sum_i \hat w_s(m_i)
= \frac
{\widehat W^B_{bb} \sum_i \frac{s_i}{g_i} - \widehat W^B_{sb} \sum_i \frac{b_i}{g_i}}
{\widehat W^B_{ss} \widehat W^B_{bb} - [\widehat W^B_{sb}]^2} \\
= N
\left(
  \sum_{ik} \frac{s_i b_k^2 - b_i s_k b_k}{g_i g_k^2}
\right) \Bigg/ \left(
  \sum_{ik} \frac{s_i^2 b_k^2  - s_i b_i s_k b_k}{g_i^2 g_k^2}
\right).
\end{gather}
In the last step, \eq{wxyb} was inserted and the products of sums expanded. We note that the denominators of the two terms in braces differ by a factor $g_i$ and convert the first term:
\begin{gather*}
\sum_{ik} \frac{s_i b_k^2 - b_i s_k b_k}{g_i g_k^2} = \sum_{ik} \frac{g_i \, (s_i b_k^2 - b_i s_k b_k)}{g^2_i g_k^2} \\
= \sum_{ik} \frac{\hat z s_i^2 b_k^2 + (1-\hat z) s_i b_i b_k^2 - \hat z s_i b_i s_k b_k - (1-\hat z) b_i^2 s_k b_k}{g_k^2 g^2_i} \\
= \sum_{ik} \frac{\hat z (s_i^2 b_k^2 - s_i b_i s_k b_k)}{g_k^2 g^2_i}.
\end{gather*}
We used $\sum_{ik} s_i b_i b_k^2 = \sum_{ik} b_i^2 s_k b_k$ in the last step. Finally, by inserting this intermediate result back we  find the desired result,
\begin{multline}
\sum_i \hat w_s^B(m_i)
= \\
N
\left(
  \sum_{ik} \frac{\hat z (s_i^2 b_k^2 - s_i b_i s_k b_k)}{g_k^2 g^2_i}
\right) \Bigg/
\left(
  \sum_{ik} \frac{s_i^2 b_k^2  - s_i b_i s_k b_k}{g_i^2 g_k^2}
\right) \\ = N \hat z.
\end{multline}

\section{Proof that sum of all component weights is unity}
\label{sec:sum_weights_unity}

When $I(m)$ is a linear combination of the \pdfs,
\begin{equation}
    I(m) = \displaystyle\sum_{k=0}^{n} a_k g_k(m),
\end{equation}
then the normalisation of the $g_k(m)$ implies that
\begin{multline}
       1 = \int \dd{m} g_k(m) = \int \dd{m} g_k(m) \frac{I(m)}{I(m)} \\
      = \displaystyle\sum_{l=0}^{n} a_l \int \dd{m} \frac{g_k(m)g_l(m)}{I(m)}
      = \displaystyle\sum_{l=0}^{n} a_l W_{kl}.
\end{multline}

Inverting this matrix equation, it follows that \mbox{$a_l = \sum_{k=0}^n A_{kl}$} and thus,
\begin{equation}
    \displaystyle\sum_{k=0}^{n} w_k(m) =  \displaystyle\sum_{k=0}^{n}  \displaystyle\sum_{l=0}^{n} \frac{A_{kl}g_l(m)}{I(m)}
    = \frac{1}{I(m)}\displaystyle\sum_{l=0}^{n} a_l g_l(m) = 1.
\end{equation}

\section{The variance function which minimises the variance of $\hat{z}$}
\label{sec:app:min_var_zhat}

Recall that an estimate for the fraction $z_k$ is given by
\begin{equation}
    \label{eq:app_zhat}
    \hat{z}_k = \frac{D}{N} \displaystyle\sum_{i=1}^{N} \frac{w_k(m_i)}{\epsilon(m_i,t_i)}.
\end{equation}
Given that $\ex[\hat{z}_k] = z_k$ then
\begin{equation}
	\ex\!\left[ \frac{w_k(m_i)}{\epsilon(m_i,t_i)} \right] = \frac{z_k}{D}.
\end{equation}
Here the normalisation $D$ is an unknown constant and for the following it is sufficient to simply assume that $D$ exists. As an aside, if one assumes a functional form of $I(m)$ which provides weights which sum to unity (Appendix~\ref{sec:exp_var_sum_w} shows that any linear combination will satisfy this requirement) and noting that the estimates $\hat{z}_k$ also have to sum to unity, then $D$ can be estimated from the data using the harmonic average of the efficiencies,
\begin{equation}
	\hat{D} = \left( \frac{1}{N} \displaystyle\sum_{i=1}^{N} \frac{1}{\epsilon(m_i,t_i)} \right)^{-1}.
\end{equation}
Assuming simply that $D$ exists, then following from Eq.~\ref{eq:app_zhat}, the variance of $\hat{z}_k$ is
\begin{multline}
	\var( \hat{z}_k) = \ex[ \hat{z}_k^2 ] - \ex [ \hat{z}_k ] ^2 \\
		= \frac{D^2}{N^2} \ex\!\left [ \displaystyle\sum_{i,j=1}^{N} \frac{w_k(m_i)w_k(m_j)}{\epsilon(m_i,t_i)\epsilon(m_j,t_j)} \right] - z_k^2  \\
		= \frac{D^2}{N^2} \left[  \displaystyle\sum_{i \neq j}^{N} \ex\!\left[ \frac{w_k(m_i)w_k(m_j)}{\epsilon(m_i,t_i)\epsilon(m_j,t_j)} \right] + \displaystyle\sum_{i= j}^{N} \ex\!\left[ \frac{w^2_k(m_i)}{\epsilon^2(m_i,t_i)} \right] \right]- z_k^2  \\
		= \frac{D^2}{N^2} \left[  N(N-1)\frac{z_k^2}{D^2} + N \ex\!\left[ \frac{w^2_k(m)}{\epsilon^2(m,t)} \right] \right]- z_k^2  \\
		= \frac{1}{N}\left[ D^2 \ex\!\left[ \frac{w^2_k(m)}{\epsilon^2(m,t)} \right] - z_k^2 \right].
\end{multline}
If the weight $I(m)$ is to be such that the variance of $\hat{z}_k$ is minimal it then follows that the expectation value $\ex[ w^2_k(m)/\epsilon^2(m,t)]$ is minimal. The minimisation has to incorporate the constraints that the integrals of $w_k(m)g_l(m)$ are either zero or one, which is done by Lagrange multipliers, $2\lambda_{l}$. The extremum condition becomes
\begin{equation}
  \int \dd{m} \dd{t} \rho(m,t) \left[\frac{w^2_k(m)}{\epsilon^2(m,t)} - \displaystyle\sum_{l=0}^{n} 2 \lambda_l w_k(m)g_l(m) \right] \overset{!}{=}` \text{min}.
\end{equation}
Here only $\rho(m,t)$ and $\epsilon(m,t)$ depend on $t$. Encompassing the $t$-integral by introducing
\begin{equation}
    q(m) = \int \dd{t} \frac{\rho(m,t)}{\epsilon^2(m,t)}
\end{equation}
and using the extremum condition, which requires that any variations $\delta w_k(m)$, with $\delta w^2_k(m) = 2w(m)\delta w(m)$, lead to zero variation of the remaining $m$ integral, one finds
\begin{equation}
    \int \dd{m} \, 2\,\delta w_k(m) \left[ w_k(m)q(m) - \displaystyle\sum_{l=0}^n \lambda_l g_l(m) \right] = 0.
\end{equation}
This is true under any variations $\delta w_k(m)$ provided the term in square brackets zero. This implies that the functional form of the weight functions is
\begin{equation}
    w_k(m) = \displaystyle\sum_{l=0}^n \frac{\lambda_l g_l(m)}{q(m)},
\end{equation}
which in turn means that the optimal variance weight function is given by
\begin{equation}
    I(m) = q(m) = \int \dd{t} \frac{\rho(m,t)}{\epsilon^2(m,t)}.
\end{equation}

\section{The variance function which means $\hat{z}_k$ are their maximum likelihood estimates.}
\label{sec:app:max_like_zhat}

Consider an Extended Maximum Likelihood fit of the yields, $N_k$, for each component of the data model. The Maximum Likelihood (ML) estimates, $\hat{N}_k$ are obtained by minimising
\begin{equation}
    \mathcal{L} = \displaystyle\sum_{l=0}^n N_k - \displaystyle\sum_{i=1}^N \frac{1}{\epsilon(m_i,t_i)} \ln \left[ \displaystyle\sum_{l=0}^n N_l g_l(m) \right].
\end{equation}
The requirement of a stationary point $\partial \mathcal{L}/ \partial \hat{N}_k = 0$ leads to
\begin{equation}
    1 = \displaystyle\sum_{i=1}^N \frac{1}{\epsilon(m_i,t_i)} \frac{g_k(m_i)}{\sum_l \hat{N}_l g_l(m_i)} .
\end{equation}
Inserting the estimates $\hat{z}_k = \hat{N}_k D/N$ means that
\begin{equation}
    \label{eq:app_ND}
    \frac{N}{D} = \displaystyle\sum_{i=1}^N \frac{1}{\epsilon(m_i,t_i)} \frac{g_k(m_i)}{\sum_l \hat{z}_l g_l(m_i)} \; \forall \; k.
\end{equation}
The solution for this system of non-linear equations requires that the right-hand-side is the same for all $k$, namely $N/D$. Noticing here the similarity with Eq.~\ref{eq:app_zhat}, one can choose $I(m)$ such that the sum in Eq.~\ref{eq:app_ND} becomes $N/D$. In this case one finds that
\begin{equation}
    \hat{z}_k = \displaystyle\sum_{l=0}^n A_{kl} = a_k
\end{equation}
and therefore
\begin{equation}
    I(m) = \displaystyle\sum_{l=0}^n \hat{z}_l g_l(m).
\end{equation}

\section{Choice of signal \pdf for COWs when the signal factorises.}
\label{sec:app:sig_density}

Quite often the signal shape $g_0(m)$ is a non-trivial function usually containing a number of nuisance parameters. It is interesting to investigate what happens to the corresponding weight function when the choice of function used in the determination of the weight function does not match the true signal density. Now we are making a distinction between the true \pdfs, $g_k(m)$ and the \pdfs used to determine the $W_{kl}$, $G_k(m)$. In this case we are discussing the signal so will assume that $G_0(m)\neq g_0(m)$ and $G_k(m) = g_k(m)$ for $k>0$. In this case we can write the expected number of signal in a bin of width $\Delta t$ in the control variable distribution as
\begin{align}
	\ex[ w_0 ] &= \int_{\Delta t} \dd{t} \int \dd{m} \rho(m,t) w_0(m) \nonumber \\
    & = \int_{\Delta t} \dd{t} \int \dd{m} \displaystyle\sum_{k=0}^n z_k g_k(m)h_k(t) \displaystyle\sum_{l=0}^n A_{0l} \frac{G_l(m)}{I(m)} \nonumber \\
    &= \displaystyle\sum_{k=0}^n z_k \int_{\Delta t} \dd{t} h_k(t) \displaystyle\sum_{l=0}^n A_{0l} \int \dd{m} \frac{G_l(m)g_k(m)}{I(m)}.
\end{align}
For $k>0$ the $m$-integral is equal to $W_{lk}$. However, for $k=0$, it is not because $G_0(m)\neq g_0(m)$. Explicitly splitting the sum over $k$ into these two parts gives
\begin{multline}
	\ex[ w_0 ] = z_0 \int_{\Delta t} \dd{t} h_0(t) \displaystyle\sum_{l=0}^n A_{0l} \int \dd{m} \frac{G_l(m)g_0(m)}{I(m)} \\
    + \displaystyle\sum_{k=1}^n z_k \int_{\Delta t} \dd{t} h_k(t) \underbrace{\displaystyle\sum_{l=0}^n A_{0l} W_{lk}}_{\delta_{0k}}.
\end{multline}

Since the $A$ and $W$ matrices are the inverse of each other, the last sum reduces to $\delta_{0k}$, and therefore the second term does not contribute to the expectation value and vanishes. This leaves,
\begin{equation}
	\ex[ w_0 ] = z_0 \left[ \displaystyle\sum_{l=0}^n A_{0l} \int \dd{m} \frac{G_l(m)g_0(m)}{I(m)} \right] \int_{\Delta t} \dd{t} h_0(t).
\end{equation}
This shows that, since the term in the square bracket is a constant, in order to determine the shape $h_0(t)$ both the $I(m)$ and $G_0(m)$ functions of the COW to project out the signal are arbitrary.

\section{Sample estimate for variance of the quasi-score vector}
\label{app:sampleestimates}
Below we give the sample estimate for $\bm C_{\bm S} = \ex[\bm{S}\bm{S}^T\bigr]$ in Eq.~\ref{eq:sandwichexpectation}. We obtain
\begin{align*}
    \widehat{\ex}\left[\frac{\partial\lnL}{\partial N_x} \frac{\partial\lnL}{\partial N_y}\right] &=
    \sum_i\frac{\hat g_x(m_i) \hat g_y(m_i)}{\bigl(\hat N_s \hat g_s(m_i) +\hat N_b \hat g_b(m_i)\bigr)^2} \\
    \widehat{\ex}\left[\frac{\partial\lnL}{\partial N_x}\frac{\partial\lnL}{\partial \phi_k}\right] &=
    \sum_i\frac{\hat g_x(m_i)\bigl(\hat N_s \frac{\partial g_s(m_i)}{\partial\phi_k} + \hat N_b \frac{\partial g_b(m_i)}{\partial\phi_k}\bigr)|_{\hat{\bm{\phi}}}}{\bigl(\hat N_s \hat g_s(m_i) + \hat N_b \hat g_b(m_i)\bigr)^2} \\
    \widehat{\ex}\left[\frac{\partial\lnL}{\partial N_x} \psi_{(uv)}\right] &=
    \sum_i\frac{\hat g_x(m_i)\hat g_u(m_i) \hat g_v(m_i)}{\bigl(\hat N_s \hat g_s(m_i) + \hat N_b \hat g_b(m_i)\bigr)^3} \\
    \widehat{\ex}\left[\frac{\partial\lnL}{\partial N_x}\xi_k\right] &=
    \sum_i \frac{\hat w_s(m_i) \hat g_x(m_i)}{\hat N_s \hat g_s(m_i) + \hat N_b \hat g_b(m_i)}\frac{\partial \ln h_s(t_i)}{\partial\theta_k}\biggr|_{\hat{\bm\theta}} \\
    \widehat{\ex}\left[\frac{\partial\lnL}{\partial \phi_k}\frac{\partial\lnL}{\partial \phi_\ell}\right] &=
    \sum_i\frac{(\hat N_s \frac{\partial g_s(m_i)}{\partial\phi_k} + \hat N_b \frac{\partial g_b(m_i)}{\partial\phi_k})|_{\hat{\bm \phi}}}{\hat N_s \hat g_s(m_i) + \hat N_b \hat g_b(m_i)} \\
    & \hphantom{=\sum}\times \frac{(\hat N_s \frac{\partial g_s(m_i)}{\partial\phi_\ell} + \hat N_b \frac{\partial g_b(m_i))}{\partial\phi_\ell})|_{\hat{\bm \phi}}}{\hat N_s \hat g_s(m_i) + \hat N_b \hat g_b(m_i)} \\
    \widehat{\ex}\left[\frac{\partial\lnL}{\partial\phi_k}\psi_{(xy)}\right] &=
    \sum_i\frac{\hat g_x(m_i) \hat g_y(m_i)}{\bigl(\hat N_s \hat g_s(m_i) + \hat N_b \hat g_b(m_i)\bigr)^3} \\
    &\hphantom{=\sum}\times \Bigl(\hat N_s \frac{\partial g_s(m_i)}{\partial\phi_k} + \hat N_b \frac{\partial g_b(m_i)}{\partial\phi_k}\Bigr)\biggr|_{\hat{\bm{\phi}}}\\
    \widehat{\ex}\left[\frac{\partial\lnL}{\partial\phi_k} \xi_\ell\right] &=
    \sum_i\frac{(\hat N_s \frac{\partial g_s(m_i)}{\partial\phi_k} + \hat N_b \frac{\partial g_b(m_i)}{\partial\phi_k})|_{\hat {\bm \phi}}}{\hat N_s \hat g_s(m_i) + \hat N_b \hat g_b(m_i)}\\
    & \hphantom{=\sum}\times \hat w_s(m_i)\frac{\partial\ln h_s(t_i)}{\partial\theta_\ell}\biggr|_{\hat{\bm{\theta}}} \\
    \widehat{\ex}\left[\psi_{(xy)}\psi_{(uv)}\right] &=
    \sum_i\frac{\hat g_x(m_i) \hat g_y(m_i) \hat g_u(m_i) \hat g_v(m_i)}{\bigl(\hat N_s \hat g_s(m_i) + \hat N_b \hat g_b(m_i)\bigr)^4} \\
    \widehat{\ex}\left[\psi_{(xy)}\xi_k\right] &=
    \sum_i \frac{\hat w_s(m_i) \hat g_x(m_i) \hat g_y(m_i)}{\bigl(\hat N_s \hat g_s(m_i) + \hat N_b \hat g_b(m_i)\bigr)^2} \\
    &\hphantom{=\sum}\times \frac{\partial \ln h_s(t_i)}{\partial\theta_k}\biggr|_{\hat{\bm{\theta}}} \\
    \widehat{\ex}\left[\xi_{k}\xi_\ell\right] &=
    \sum_i \hat w_s^2(m_i) \left(\frac{\partial\ln h_s(t_i) }{\partial\theta_k}\frac{\partial\ln h_s(t_i) }{\partial\theta_\ell}\right)\biggr|_{\hat{\bm{\theta}}},
\end{align*}

where $\hat{g}_x(m_i) = g_x(m_i;\hat{\bm \phi})$, $h_s(t_i) = h_s(t_i,{\bm \theta})$, $\hat w_s(m_i) = w_s(m_i; \hat g_s, \hat g_b, \widehat W_{ss}, \widehat W_{sb}, \widehat W_{bb})$, $x,y \in \{s, b\}$, $(xy),(uv)$ each iterate over $\{ ss, sb, bb \}$, and $k,l$ index the shape parameters of $\bm \phi$ or $\bm \theta$.

\section{Variance of a sum of weights}\label{sec:exp_var_sum_w}

We compute the variance of a sum of independently and identically distributed weights, $T = \sum_i^{n} w_i$, where the sample size $n$ is a Poisson-distributed number. The latter changes the computation of the variance of $T$. We follow the derivation in Ref.~\cite{Benjamin:2014}; the key insight is that the sampling of $n$ is independent of the sampling of the $w_i$.

The variance of $T$ is $\var(T) = \ex[T^2] - \ex[T]^2$, so we need the respective expectations. The expectation of $T$ is
\begin{equation}
\ex[T] = \ex_n[\ex_w[T]] = \ex_n\left[\sum_i^n \ex[w]\right] = \ex[n] \, \ex[w],
\end{equation}
where $\ex_n$ is an expectation taken with respect to $n$ only, likewise for $\ex_w$. The expectation of $T^2$ is
\begin{multline}
\ex[T^2] = \ex_n[\ex_w[T^2]] = \ex_n\left[\var_w(T) + \ex_w[T]^2 \right] \\
= \ex_n\left[\sum_i^n \, \var(w) + n^2 \, \ex[w]^2 \right] \\
= \ex[n] \, \var(w) + \ex[n^2] \, \ex[w]^2.
\end{multline}
Here we used that the variance of a sum of independent random variables is equal to the sum of their variances. The variance of $T$ then is
\begin{multline}
\var(T) = \ex[n] \, \var(w) + \ex[n^2] \, \ex[w]^2 - \ex[n]^2 \, \ex[w]^2 \\
= \ex[n] \, \var(w) + \var(n) \, \ex[w]^2.
\end{multline}
With $\var(n) = \ex[n]$ for a Poisson distribution, the variance reduces to
\begin{equation}
\var(T) = \ex[n] \, (\var(w) + \ex[w]^2) = \ex[n] \, \ex[w^2].
\end{equation}
An unbiased estimate of this is given by
\begin{equation}
\widehat{\var}(T) = n \times \frac1n \sum_i w_i^2 = \sum_i w_i^2.
\end{equation}

\bibliography{main}

\end{document}